\documentclass[11pt]{article}
\def\DESepsf(#1 width #2){\epsfxsize=#2 \epsfbox{#1}}
\usepackage[dvips]{graphicx}
\usepackage{latexsym}
\usepackage{amssymb}
\usepackage{times}
\usepackage{mathptm}
\def\th13 {\theta_{13}}

\def\bc {\begin{center}}
\def\ec {\end{center}}
\def\be {\begin{equation}}
\def\bea {\begin{eqnarray}}
\def\ee {\end{equation}}
\def\eea {\end{eqnarray}}

\def\lapp{\mathrel{\rlap{\raise.5ex\hbox{$<$}}
                    {\lower.5ex\hbox{$\sim$}}}}
\def\gapp{\mathrel{\rlap{\raise.5ex\hbox{$>$}}
                    {\lower.5ex\hbox{$\sim$}}}}

\begin{document}

\begin{flushright} 
HRI-P-08-12-004 
\end{flushright} 

\bc{\bf \Large A comparison of the sensitivities of  the
parameters}
\vskip 5pt
\bc{\bf \Large with atmospheric neutrinos for different analysis methods}
\ec
\vskip .1in
%\author 
{\bf \large Abhijit Samanta
\footnote{E-mail address: abhijit@hri.res.in}}
%\affiliation
\vskip .20in
{\em Harish-Chandra Research Institute,
Chhatnag Road,
Jhusi,
Allahabad 211 019,
India}

{\today}
\ec

\begin{abstract}
\noindent
In the atmospheric neutrino experiments the primary problems are the 
huge uncertainties of flux, very rapid fall of flux with increase of energy,
the energy dependent wide resolutions of energy and zenith angle between true 
neutrinos  and reconstructed neutrinos. These all in together 
make the choice of binning of the data for chi-square analysis complicated.  
The large iron calorimeter has the ability to measure the 
energy and the direction of the muon with high resolution.
From the bending of the track in the magnetic field it can also
distinguish its charge. We have analyzed the atmospheric neutrino
oscillation generating events  by  Nuance and then 
considering the muons produced in the charge current interactions
as the reconstructed neutrinos. This practically takes into account the
major problem of wide resolutions. We have binned the data in three ways:
i) in the grids  of $\log E -\log L$ plane, ii) in the grids of 
$\log E -\cos\theta_{\rm zenith}$ plane, and iii) in the bins of $\log (L/E)$. 
We have performed a marginalized $\chi^2$ study over $\Delta m_{32}^2, ~\theta_{13}$
and $\theta_{23}$ for neutrinos and anti-neutrinos separately for each method
and finally compared the results. 
\end{abstract}

{\bf PACS} {14.60.Pq}

{\texttt Keywords: neutrino oscillation, atmospheric neutrino, INO}

%\maketitle
\section{Introduction}

The atmospheric neutrino anomaly was
first observed by IMB in 1986 and then confirmed 
by Kamiokande in 1988 \cite{Haines:1986yf,Hirata:1988uy}.
Finally, the neutrino oscillation was  discovered in 1998 with atmospheric neutrino
experiment \cite{Fukuda:1998mi}. The atmospheric neutrinos are produced by the 
interactions of the cosmic rays with the atmosphere. At the neutrino energies above a few GeV, 
the effect of geo-magnetic field on the cosmic rays
is negligible and then 
the atmospheric
neutrino flux can be predicted to be up-down symmetric. 
 The flight lengths for up and down going neutrinos are very
different. The atmospheric neutrino experiments exploit these features to
study the neutrino oscillation.

The atmospheric neutrinos are also equally important in the precision era of 
neutrino physics. The main thrust is now on the precise measurements of  oscillation 
parameters.  This helps to identify the right track to understand the underlying principle that gives the neutrino masses 
and their mixing. In the recent years, the studies of neutrinos has become a popular tool
to probe the physics
beyond the standard model.
In the standard oscillation picture, there are six parameters.  
The present 1$\sigma$, 2$\sigma$ and 3$\sigma$ confidence level ranges
from global $3\nu$ oscillation analysis (2008)
\footnote{The CP-violating phase $\delta_{CP}$ is still unconstrained.}
\cite{Fogli:2008ig} are tabulated in
table \ref{t:global-fit}.
%===========================================================================
\begin{table}[htb]
%\centering
%%%%%\resizebox{\textwidth}{!}{
%%%%%\begin{ruledtabular}
\begin{tabular}{cccccc}
\hline
Parameter & $\Delta m_{21}^2/10^{-5}\mathrm{\ eV}^2$ & $\sin^2\theta_{12}$ & $\sin^2\theta_{13}$ & $\sin^2\theta_{23}$ &
$|\Delta m_{31}^2|/10^{-3}\mathrm{\ eV}^2$ \\[4pt]
\hline%---------------------------------------------------------------------
Best fit        &     7.67     &  0.312          &  0.016          &  0.466          &  2.39 \\
$1\sigma$ range & 7.48~--~7.83 & 0.294~--~0.331  & 0.006~--~0.026  & 0.408~--~0.539  & 2.31~--~2.50 \\
$2\sigma$ range & 7.31~--~8.01 & 0.278~--~0.352  & $<0.036$        & 0.366~--~0.602  & 2.19~--~2.66 \\
$3\sigma$ range & 7.14~--~8.19 & 0.263~--~0.375  & $<0.046$        & 0.331~--~0.644  & 2.06~--~2.81 \\
\hline
\end{tabular}
%%%%%\end{ruledtabular}
%%%%}%end of resizebox
%\vspace*{.2cm}
\caption{\sf \small Global 3$\nu$ oscillation analysis (2008)}
\label{t:global-fit}
\end{table}

This spectacular achievement is very stimulating to uncover the facts which are 
still missing. To determine the mass ordering  
(sign of $\Delta m_{32}^2$) \footnote{$\Delta m^2_{32}=m_3^2-m_2^2$.}, 
the values of $\theta_{13}$ and  $\delta_{CP}$ with good precision, the octant of $\theta_{23}$ 
with atmospheric neutrinos as well as neutrinos from artificial 
beams, 
there are many ongoing and planned experiments:  INO \cite{Arumugam:2005nt}, 
UNO \cite{Jung:1999jq}, T2K \cite{Itow:2001ee}, NOvA \cite{Ayres:2004js}, 
 Hyper-Kamiokande \cite{Nakamura:2003hk} and many others.
In the current few years, a large fraction of effort in particle physics 
research has gone to study the physics potential of these detectors 
\cite{GonzalezGarcia:2007ib}.  
The current research activity \cite{Choubey:2005zy,
Gandhi:2007td,
Indumathi:2004kd, Petcov:2005rv, Indumathi:2006gr, Agarwalla:2005we, Agarwalla:2006vf,
Agarwalla:2008gf, Datta:2003dg, 
inopaper} shows the uniqueness in physics potential of 
the large magnetized Iron CALorimeter (ICAL) detector at the
India-based Neutrino Observatory (INO). It should be noted here that
its position at PUSHEP has a special feature. It gives the {\it magic baseline from CERN} 
for beam experiments, which provides the oscillation probabilities relatively
insensitive to the yet unconstrained CP phase compared to all other baselines 
and permits to make the
precise measurements of the masses and their mixing avoiding the 
degeneracy issues \cite{Agarwalla:2005we}.
On the other hand, ICAL can detect $\nu_\mu$ and $\bar\nu_\mu$ separately
using the magnetic field for charge current events.
The oscillation study with atmospheric neutrinos is the primary goal of ICAL at INO. 
Before going into the detailed techniques of the analysis methods, we will first discuss the 
basic nature of atmospheric neutrino oscillation and the detection characteristics
of ICAL detector. 

\subsection{ The atmospheric neutrino oscillation and the ICAL detector}
The present atmospheric neutrino data from the pioneering Super Kamiokande (SK) experiment are well explained 
by two flavor 
oscillation \cite{Ashie:2005ik, Ashie:2004mr}.
However, one expects the reflection of  $\nu_\mu\rightarrow\nu_e$ 
oscillation in data for standard 3-flavor framework in the data if $\theta_{13}$ is nonzero.
Neglecting the $\Delta m_{21}^2$ term the oscillation probability
can be expressed as:
\begin{eqnarray}
%\mbox{P}(\nu_e \rightarrow \nu_e) &=& 1- \sin^2 2\theta_{13}
%      \sin^2 \left(\frac{1.27 \Delta m^2 L}{E}\right) \nonumber \\
\mbox{P}(\nu_\mu \rightarrow \nu_e) &=& \mbox{P}(\nu_e \rightarrow \nu_\mu) \nonumber \\
&=& \sin^2 \theta_{23} \sin^2 2 \theta_{13}
      \sin^2 \left(\frac{1.27 \Delta m^2 L}{E}\right) \nonumber \\
\mbox{P}(\nu_\mu \rightarrow \nu_\mu) &=& 1 \nonumber \\
&& - 4 \cos^2 \theta_{13}
\sin^2 \theta_{23} ( 1-\cos^2 \theta_{13} \sin^2 \theta_{23}) \nonumber \\
&&\times 
\sin^2 \left(\frac{1.27 \Delta m^2 L}{E}\right)
\label{eqn:oscillation-vacuum}
\end{eqnarray}
These oscillation probabilities are derived for vacuum.  Since  it
involves electron neutrino, the oscillation will be modulated by the matter
effect \cite{Mikheev:1986gs,Wolfenstein:1977ue}. Then,
\begin{eqnarray}
\mbox{P}(\nu_\mu \rightarrow \nu_e) &=& \mbox{P}(\nu_e \rightarrow \nu_\mu) \nonumber \\
&=& \sin^2 \theta_{23} \sin^2 2 \theta_{13}^M
      \sin^2 \left(\frac{1.27 \Delta {m^2}_{M} L}{E}\right). 
\label{e:oscillation-matter}
\end{eqnarray}
The symbol `M' denotes effective parameters in matter. 
The effective mixing angle is
\begin{eqnarray}
\sin^2 2\theta_{13}^M &=& 
      \frac{\sin^2 2\theta_{13}}{(\cos2\theta_{13}-A_{CC}/\Delta m^2)^2 
      + \sin^22\theta_{13}} 
\label{e:effective-mixing}
\end{eqnarray}
and
\begin{eqnarray}
 \Delta {m^2}_{M} &=& \sqrt{\left(\Delta m^2\cos2\theta_{13}-A_{CC}\right )^2 +
\left( \Delta m^2 \sin2\theta_{13}\right)^2}
\end{eqnarray}
with 
\begin{eqnarray}
A_{CC} &=& 2\sqrt{2}G_FN_e E, 
\end{eqnarray}
where \(G_F\) is the Fermi constant, \(N_e\) is the electron density
of the medium and \(E\) is neutrino energy \cite{Giunti:1997fx}.
The matter potential term \(A_{CC}\) has the same absolute value, but opposite
sign for neutrinos and anti-neutrinos.
The Mikheyev-Smirnov-Wolfenstein (MSW) resonance
occurs 
%in $\mbox{P}(\nu_\mu \rightarrow \nu_e)$ or  $\mbox{P}(\nu_e \rightarrow \nu_\mu)$ 
when neutrino %$\nu_e$ ($\bar\nu_e$)
passes through the matter
(see eq. \ref{e:effective-mixing}). 
It happens for Normal Hierarchy (NH)
with neutrinos and for Inverted Hierarchy (IH) with anti-neutrinos. 
The resonance energy corresponding to a baseline can be seen 
in  \cite{Samanta:2006sj}.

The muon neutrino (anti-neutrino) produces $\mu^-$ ($\mu^+$) in Charge Current
(CC) weak interactions. The magnetized ICAL can 
distinguish $\mu^+$ and $\mu^-$ with the magnetic field.  
The energy ($E$) and zenith angle ($\theta_{\rm zenith}$) or baseline 
($L$) resolutions of the muons are  
very high at ICAL \cite{Arumugam:2005nt}.
The hadron energy can also be measured at ICAL. However, its resolution  is  very poor and 
strongly depends on thickness of the iron layers.

The atmospheric neutrinos 
are expected to be very useful in precision studies for  
its very wide energy range (MeV $-$ few hundred GeV) and wide baseline range
(few km $-12950$ km). It gives both neutrino and anti-neutrino, which
behave oppositely with matter. This helps to detect the 
sign( $\Delta m_{32}^2$), the value of $\theta_{13}$ as well as the octant of 
$\theta_{23}$. 
One can exploit this feature to measure the precision of  
these parameters and the mass ordering at the magnetized 
ICAL detector  at INO. It should be noted here that the non-magnetized
detectors, like water Cherenkov detector, can also contribute in this 
study since the cross section,  the $y(=(E_\nu-E_{\rm lepton})/E_\nu)$ 
dependence of the cross section are  different for $\nu$ and $\bar\nu$. 
The water detectors may also be able to distinguish statistically $\nu_\mu$ 
and $\bar\nu_\mu$ due to different capture rates and lifetimes of the charged 
muons in water.

However, one of the crucial problems in neutrino physics experiments is the 
wide resolutions of $E$ and $L$ between true neutrinos and 
reconstructed neutrinos, which smears the oscillation effect to some
significant extent. This arises mainly due to interaction kinematics.
The un-observable product particles, un-measurable momentum of recoiled nucleus
are the main sources of this huge uncertainty in reconstructed neutrino momentum. 
These are strongly neutrino energy dependent.

Due to the above complications, the method of extraction of the results from the data
is not straightforward.  The results depend crucially on
the way of the analysis and particularly on the type of binning of the data. 
This fact is well-known from the analysis of atmospheric neutrino data of
SK experiment \cite{Ashie:2005ik, Ashie:2004mr}.
In this paper we consider the reconstructed energy and the direction of 
an event only from the muon 
generating it by the neutrino event generator Nuance-v3\cite{Casper:2002sd}.
{The addition of hadrons to the muon, which might increase the 
reconstructed neutrino energy resolution, is not considered here for 
conservative estimation of the sensitivity. It would be realistic in case 
of GEANT-based studies since the number of hits produced by the 
hadron shower  strongly depends on the iron thickness.
However, INO can also detect the neutral current events. Though it is
expected that these will not have any directional information, the energy
dependency of the averaged oscillation  over all directions can also contribute 
to the total $\chi^2$ separately in the sensitivity studies. 
}
Here we have studied the atmospheric neutrino oscillation by binning the data
in different ways and finally compared the results. These are discussed 
in the next sections.

\section{The $\chi^2$ analysis}
Now we will describe a general expression for $\chi^2$, 
the method for generation of the theoretical data, the estimated systematic uncertainties, 
and finally the ways of binning of the data. 
The number of events falls very rapidly with the increase of energy and 
the statistics is very poor 
at high energy. However,  the contribution to the sensitivities of the oscillation 
parameters is significant from these high energy events.
To incorporate  these events at high energy, the $\chi^2$ value is calculated 
according to Poisson probability distribution. 
For all types of binning, we define a general  expression of $\chi^2$ as
 \begin{eqnarray}
 \chi^2 &=& \sum_{I=1}^{N} \left[ 2 \left\{ N^{p}_{I} 
  - N^{o}_{I} \right\} - 2 N^{o}_{I} \ln \left( \frac{N^p_I}
{N^o_I}\right)  \right]
  + \sum_{k=1}^{n_s} {\xi_k}^2
 \label{e:chisq}
 \end{eqnarray}
with
 \begin{eqnarray}
N^{p}_{I} &=& \sum_{i,j=n_c^{\rm low},n_E^{\rm low}}
^{n_c^{\rm high},n_E^{\rm high}}  N^{p}_{ij} \left(
 1+\sum_{k=1}^{n_s} f^k_{ij} \cdot
 \xi^k \right),\nonumber\\ {\rm and}\\
N^{o}_{I} &=& \sum_{i,j=n_c^{\rm low},n_E^{\rm low}}
^{n_c^{\rm high},n_E^{\rm high}}  N^{o}_{ij}
 \end{eqnarray}
The $N^o_{ij}$ ($N^p_{ij}$) is considered as the number of observed (predicted)
events in the $ij$th grid in the plane of $\log E - \cos\theta_{\rm zenith}$. 
Here we consider the data for 1 Mton.year exposure
of the detector.
The $f^k_{ij}$ is the systematic error
of $N^p_{ij}$ due to the $k$th uncertainty.  The ${\xi_k}$ is the pull
variable for the $k$th systematic error.
We consider $n_s=5$. 
Here we have considered 30 bins of $\log E$ and
300 bins of $\cos\theta_{\rm zenith}$ for both $N^p_{ij}$ and $N^o_{ij}$. 
However, it should be noted here that in calculation of the oscillated flux we consider 200 
bins of $\log E$ and 300 bins of $\cos\theta_{\rm zenith}$ to find the accurate 
oscillation pattern.
We consider the $E$ range $0.8-50$ GeV and $\cos\theta_{\rm zenith}$ range $-1$ to $+1$.
It should be noted here that
the energy and angular resolutions between the  muons and the neutrinos of 
the events differ significantly for  neutrinos and anti-neutrinos
due to their different ways of interactions. 

\begin{figure*}[htb]
\hspace*{-2cm}
\includegraphics[width=6.8cm,angle=270]{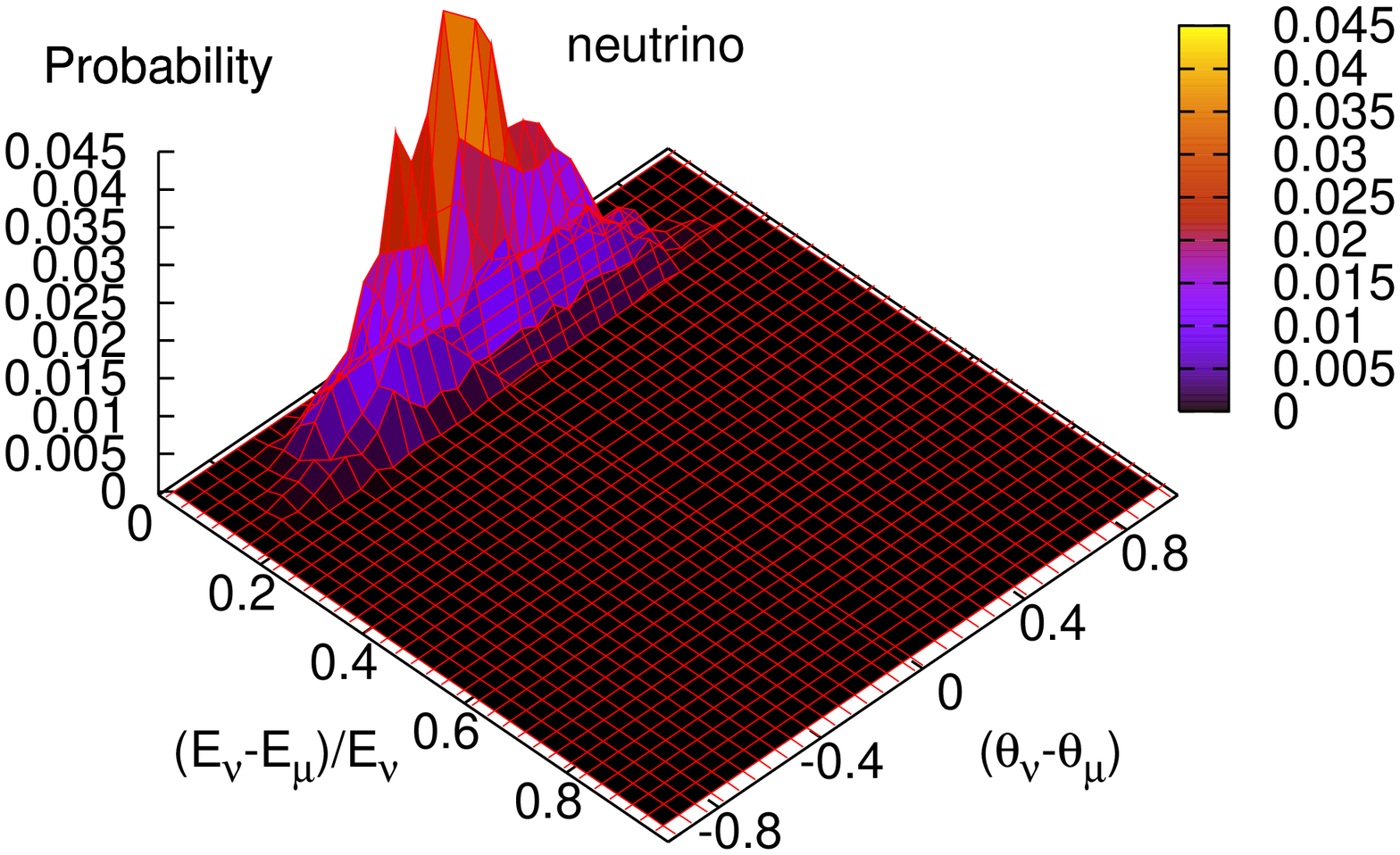}
\includegraphics[width=6.8cm,angle=270]{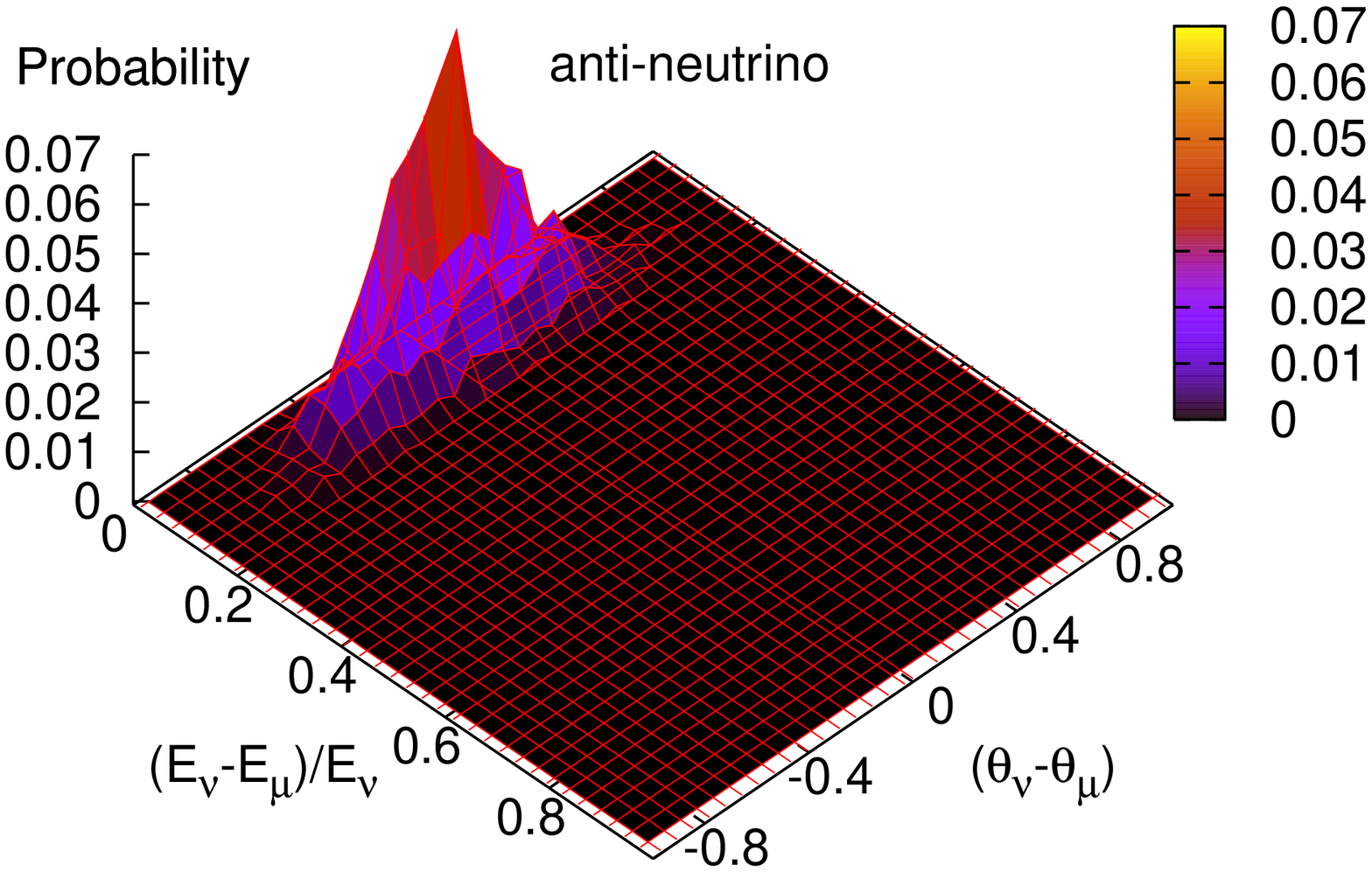}
\hspace*{-2cm}
\includegraphics[width=6.8cm,angle=270]{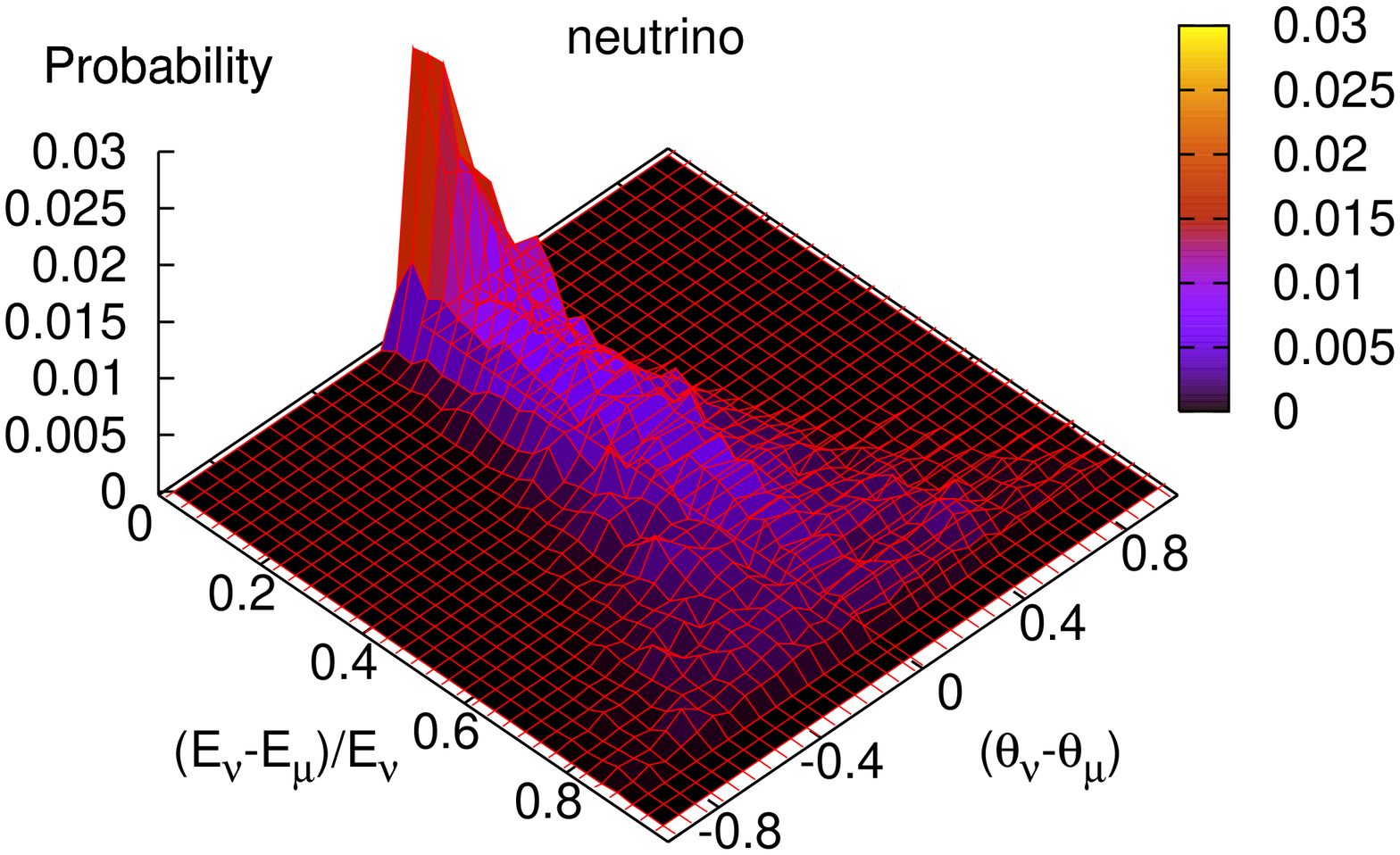}
\includegraphics[width=6.8cm,angle=270]{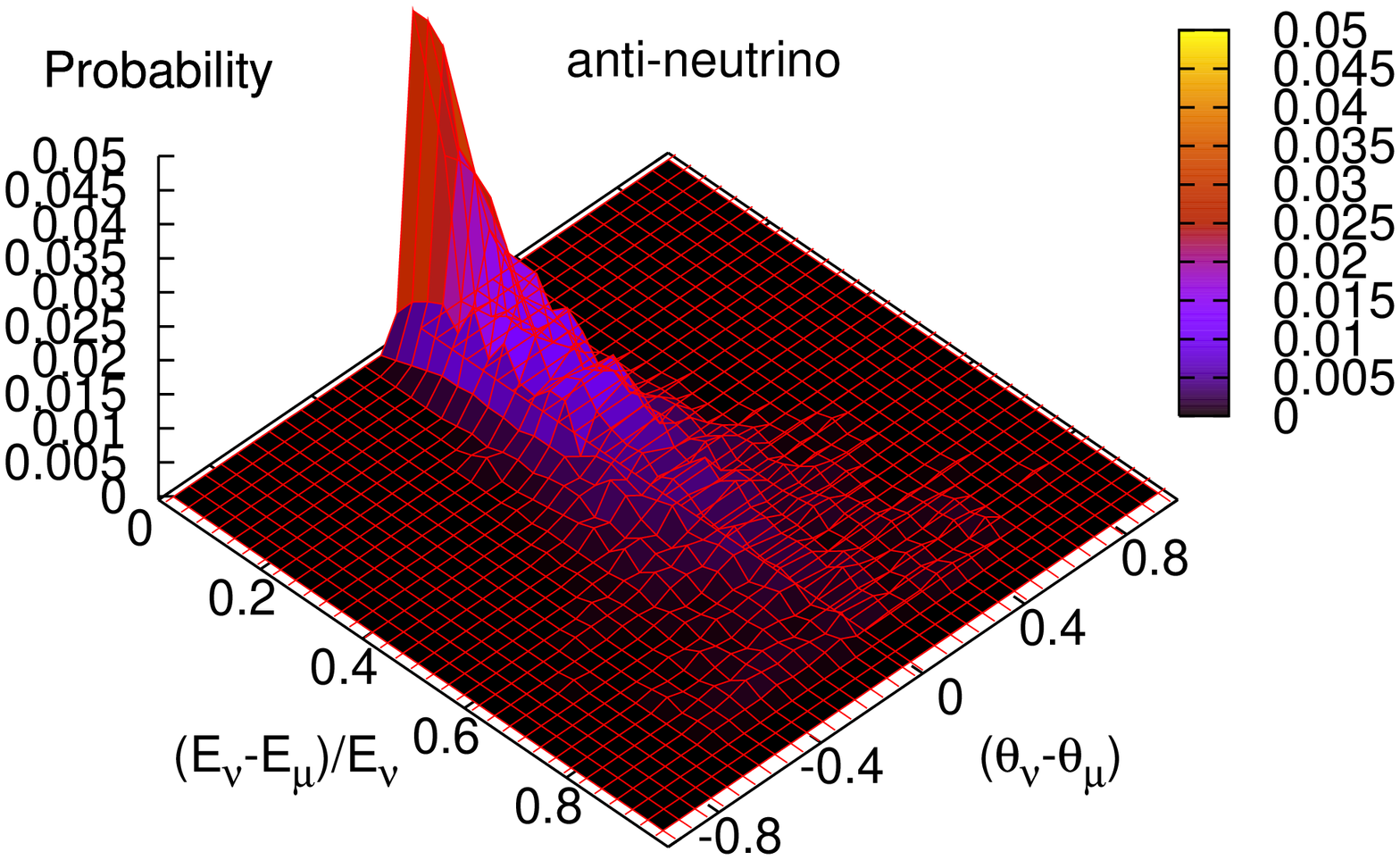}
\caption{\sf  \small The sample energy-angle correlated resolution plots for neutrino (left column) and 
anti-neutrino (right column) for the bins of $E_\nu=0.85-0.98$ GeV with
$\cos\theta_{\rm zenith}= -0.40$ to $ -0.20$ (upper row) and 
$E_\nu= 6.84 - 7.86$ GeV and $\cos\theta_{\rm zenith}=0$ to $0.20$ (lower row).
The data are obtained from the simulation of 500 MTon.year
exposure of ICAL considering no oscillation.}
\label{f:reso}
\end{figure*}

To generate the theoretical data $N^p_{ij}$ for the chi-square analysis,
we first generate 500 years un-oscillated data for 1 Mton detector.
From this data we find the energy-angle correlated resolutions (see figs. \ref{f:reso}) 
in 30 bins of energy  (in log scale) 
and 10 bins of cosine of zenith angle  ($-1$ to $+1$). For a given $E_\nu$,
we calculate the efficiency of having $E_\mu \ge$ 0.8 GeV (threshold
of the detector).
For each set of oscillation parameters, we integrate the oscillated 
atmospheric neutrino 
flux folding the total CC cross section, the exposure time, the target mass, the efficiency 
and the resolution function to obtain the predicted data in the reconstructed
$\log E-\cos\theta_{\rm zenith}$ grid
\footnote{One can do this in an another way. This is generating  the theoretical data directly
for each set of oscillation parameters. 
To ensure
that the statistical error is negligible, one needs first to generate a huge number of 
events. For example, one may generate events for 500 Mton.year exposure of the detector 
for each set of 
oscillation parameters. 
Then to obtain the theoretical data, one needs to normalize the data to  
1 Mton.year exposure of the detector dividing the events of each energy 
and zenith angle bin by 500 since the experimental data is considered 
for 1 Mton.year exposure. This would be the more 
straightforward method. 
But the marginalization study with this method is almost 
an undoable job in normal CPU. However, an exactly equivalent result is 
obtained here using the energy-angle correlated resolution function.}.
We use the CC cross section of Nuance-v3 \cite{Casper:2002sd} and the
Honda flux  of 3-dimensional scheme \cite{Honda:2004yz}.

%{\bf Systematic uncertainties:}
The atmospheric neutrino flux is not known precisely.  There are huge 
uncertainties in its estimation. We may divide them into two categories: 
I) overall uncertainties (which are
flat with respect to energy and zenith angle), and
II) tilt  uncertainties (which are  function of  energy and/or zenith angle).
These have been estimated as the following  \cite{Ashie:2005ik}:
\begin{enumerate}
\item
The energy dependence uncertainty which arises due to the uncertainty in 
spectral indices, can be expressed as:
\begin{equation}
      \Phi_{\delta_E}(E) = \Phi_0(E) \left( \frac{E}{E_0} \right)^{\delta_E}
      \approx \Phi_0(E) \left[ 1 + \delta_E \log_{10} \frac{E}{E_0} \right].
\label{e:uncer}
\end{equation}
The uncertainty of ${\delta_E}=$5\% and $E_0 = 2$~GeV
is considered.

\item
Again, the flux uncertainty as a function of  zenith angle  can be expressed as
\begin{equation}
      \Phi_{\delta_z}(\cos\theta_z)
%= \Phi_0(\cos\theta_z)
%\left( \frac{E}{E_0} \right)^\delta_z
      \approx \Phi_0(\cos\theta_z) \left[ 1 + \delta_z |\cos\theta_z| \right].
\end{equation}
The uncertainty of $\delta_z$ is considered to be $2\%$.
\item
A flux normalization uncertainty of 20\%.
\item
An over all uncertainty of 10\% in  neutrino cross section.
\item  An overall 5\% uncertainty 
for this analysis.
\end{enumerate}

We consider three types of binning:
\begin{itemize}
\item
{\bf Type I:} The events are binned in the grid of $\log E-\log L$ 
plane. We use total number of $\log E$ bins $n_E$ = 30 (0.8 $-$ 50 GeV) and the number of  $\log L$ bins  
as a function of of the energy.
We consider $n_L = 2\times 14,~2\times 18,~2\times 22,~2\times 26,$ and $~2\times 30$ for
$E= 0.8-1.2,~ 1.2-2.4,~ 2.4-3.6,~ 3.6-4.8,~ {\rm and} ~ >4.8$ GeV, respectively.
For the down-going events the  binning  is done
by replacing `$\log L$' by $`-\log L$'. 
The factor `2' is to consider both up and down going cases. 
\item
{\bf Type II:} The events are binned in the grid of $\log E-\cos\theta_{\rm zenith}$ 
plane with  exactly in the same fashion of type I. 
% .number of $E$ bins $n_E$=30   and the number of $\cos\theta_{\rm zenith}$ bins  
%$n_c = 2\times 14,~2\times 18,~2\times 22,~2\times 26,~2\times 30$ for
%$E= 0.8-1.2,~ 1.2-2.4,~ 2.4-3.6,~ 3.6-4.8,~ {\rm and} ~ >4.8$ GeV, respectively.
The only difference is that the binning is done in $\cos\theta_{\rm zenith}$ instead of $\log L$.
\item
{\bf Type III:} The events are binned in  100  $\log (L/E)$ bins and replacing 
`$\log(L/E)$' by `$-\log (L/E)$' for down-going events.  
\end{itemize}

For the up-going neutrino, $L$ is the distance traveled by the neutrino
from the detector to the source at the atmosphere. In case of down-going
neutrinos the distance traveled from the source to detector is negligible 
for getting an appreciable oscillation. However, these events help to minimize
the systematic uncertainties when considered in the $\chi^2$ analysis. The flux  
for a fixed $E$ is strongly dependent on the zenith angle. 
So, for the down-going neutrinos, we mapped the zenith angle into
$L$ considering the mirror $L$.  This  is
the same $L$ if the neutrino comes from
exactly opposite direction.
It should be noted here that
the angular error makes a much smaller error to $L$ when the tracks 
are near vertical. It increases gradually when the tracks are slanted and   
very rapidly when they are near horizontal.

For each set of oscillation parameters we calculate the $\chi^2$ in two stages.
First we used $\xi_k$ such that $\frac{\delta \chi^2}{\delta \xi_k}=0$, which 
can be obtained solving the  equations \cite{Fogli:2002pt}. Then we calculate the final $\chi^2$ with these $\xi_k$ values.
Finally, we find the minimum from these $\chi^2$ with respect to all oscillation
 parameters
\footnote{Here we consider all uncertainties as a function of reconstructed 
neutrino energy and direction. We assumed that the tilt uncertainties will
not be changed too much due to reconstruction. However, on the other hand, 
if any tilt uncertainty arises in reconstructed neutrino events from the 
reconstruction method or kinematics of scattering, 
these are then accommodated in $\chi^2$. We first incorporate all uncertainties in 
$\log E-\cos\theta_{\rm zenith}$ bins. Then we re-bin the data in the form what 
we want, {\it e.g.;}
$\log E-\log L$ bins. It should be noted that we first binned the data 
into a large number of   $\cos\theta_{\rm zenith}$ bins compared to number of 
$L$ bins to get proper binning in $\log L$.}.

\begin{figure*}[htb]
\bc
\includegraphics[width=12.0cm,angle=270]{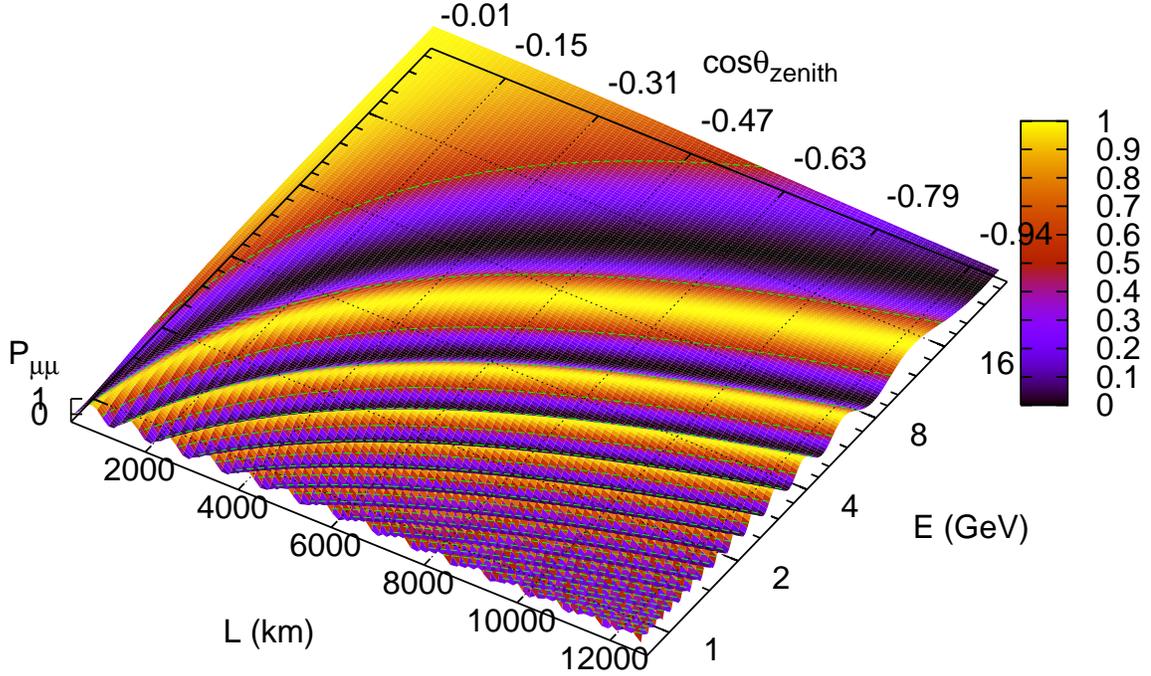}
\caption{\sf\small   The oscillation probability of $\nu_\mu\rightarrow\nu_\mu$.
We choose $\Delta m_{32}^2=-2.5\times 10^{-3}$eV$^2$,
$\theta_{23}=45^\circ$ and $\theta_{13}=0^\circ$.}
\ec
\label{f:p3d}
\end{figure*}
  
\begin{figure*}[htb]
\bc
\includegraphics[width=8.0cm,angle=0]{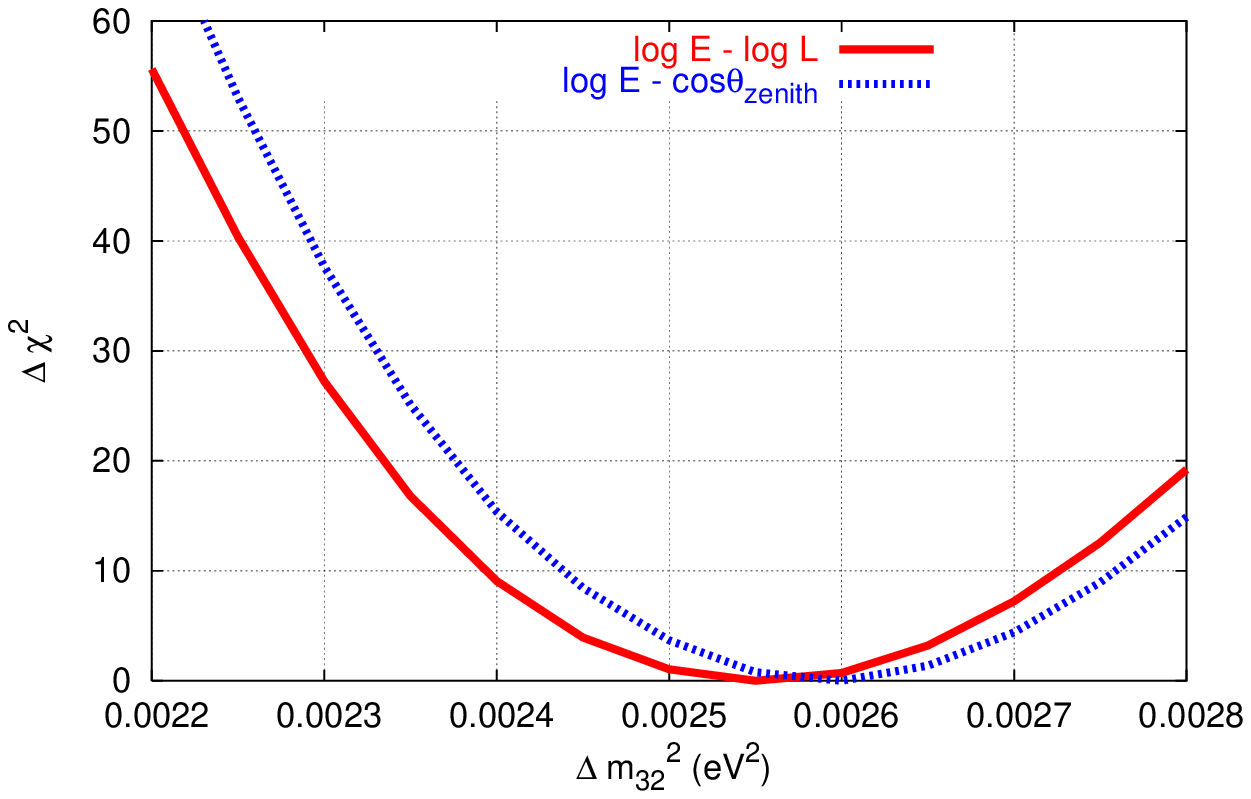}
\caption{\sf\small  The typical distribution of $\Delta \chi^2$ with 
$\Delta m^2_{32}$. We choose the input of $\Delta m_{32}^2=+2.5\times 10^{-3}$eV$^2$,
$\theta_{23}=42^\circ$ and $\theta_{13}=7.5^\circ$.}
\label{f:dcm}
\ec
\end{figure*}

\section{Result} 
In this section, we first discuss the results qualitatively  in a very 
general way 
for all analysis techniques. Then we compare the results for different
techniques. In all cases
a global scan is carried out over the  three oscillation parameters 
$\Delta m_{32}^2,~\theta_{23}$ and $\theta_{13}$ for both normal and 
inverted hierarchies with neutrinos and anti-neutrinos separately.
We have considered the range of $\Delta m_{32}^2=2.0-3.0\times 10^{-3}$eV$^2$,
$\theta_{23}=37^\circ - 54^\circ$, and $\theta_{13}=0^\circ - 12.5^\circ$.
We have fixed other parameters $\Delta m_{21}^2$ and $\theta_{12}$ at their 
best-fit values and $\delta_{\rm CP}=0$.
The 2-dimensional 68\%, 90\%, 99\% confidence level allowed parameter spaces
(APSs)  are
obtained by considering $\chi^2=\chi^2_{\rm min}+2.48,~4.83,~9.43$.
To obtain the APS in $\theta_{13}-\Delta m_{32}^2$ 
($\Delta m_{32}^2 - \theta_{23}$) plane,  
we marginalize the $\chi^2$ over $\theta_{23}~ (\theta_{13})$ over its whole 
range.

The experiment indicates that the value of $\theta_{13}$ is very small 
compared to $\theta_{23}$ \cite{Apollonio:1999ae}. 
So, the atmospheric neutrino oscillation is mainly governed by 
two flavor oscillation
$\nu_\mu~(\bar\nu_\mu)~\leftrightarrow\nu_\tau~(\bar\nu_\tau)$.
This  constrains $\sin^2 2\theta_{23}$ and $|\Delta m_{32}^2|$. 
Now, there appears a degeneracy in $\theta_{23}$ whether it is larger or smaller
than $45^\circ$ due to the $\sin^2 2\theta_{23}$ dependence of oscillation 
probability.
However, when the matter effect comes into the play,
 the effective value of $\theta_{13}$ becomes large and a  resonance occurs in 
$\nu_\mu~(\bar\nu_\mu)~\leftrightarrow\nu_e~(\bar\nu_e)$ 
oscillation. This breaks the above $\theta_{23}$ degeneracy.

\begin{figure*}[htb]
\bc
\includegraphics[width=15.0cm,height=6cm,angle=0]{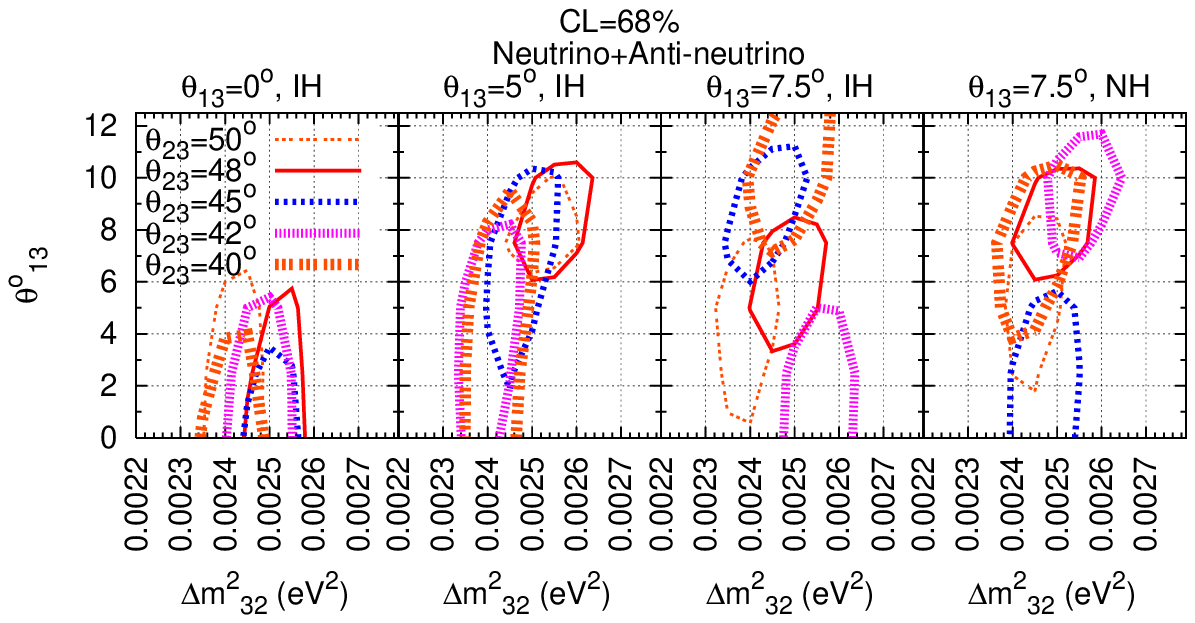}
\includegraphics[width=15.0cm,height=6cm,angle=0]{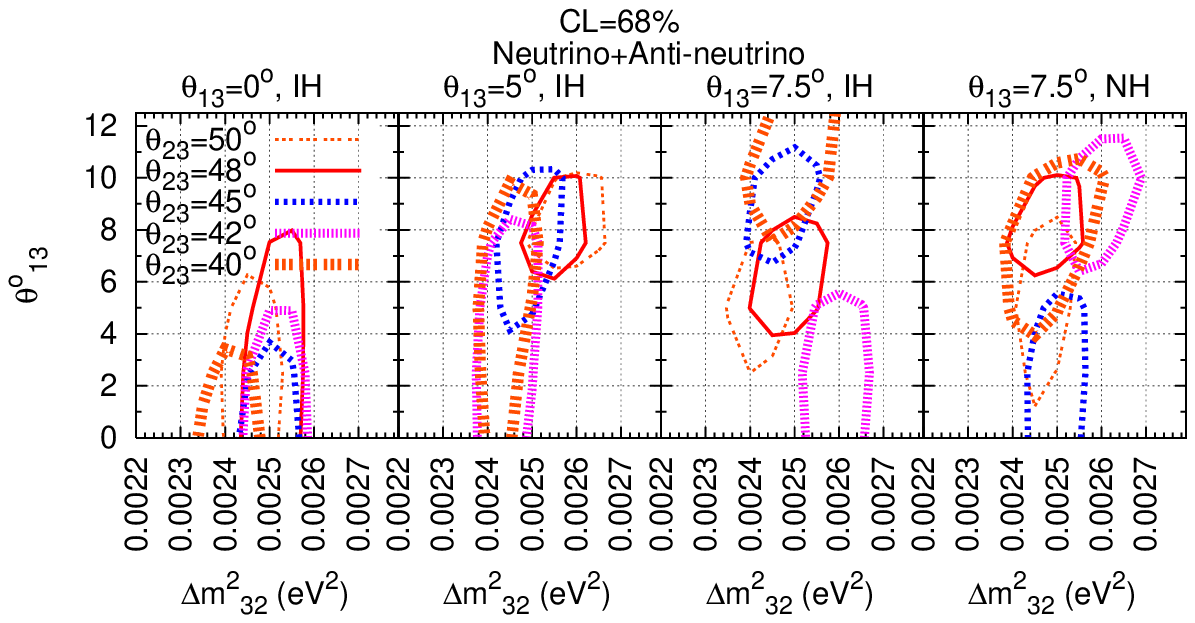}
\includegraphics[width=15.0cm,height=6cm,angle=0]{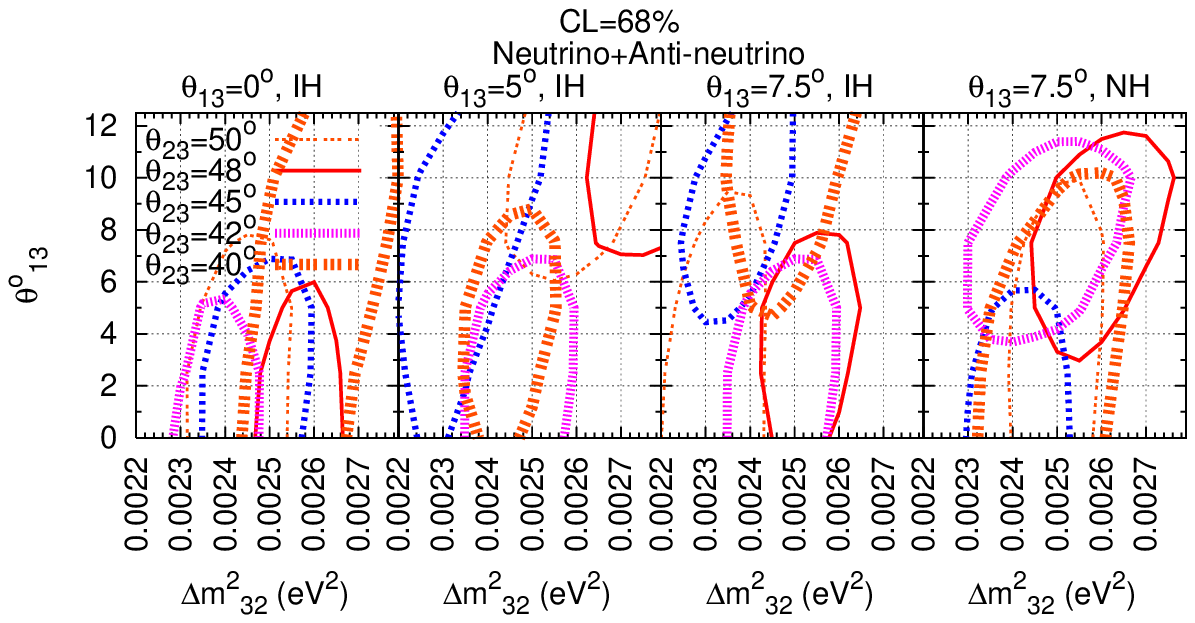}
\ec
\caption{\sf \small  
The 68\%  CL allowed regions in 
$\theta_{13}-\Delta m_{32}^2$ plane for  type I (top) type II (middle) and type III 
(bottom)  binning of the data  with the input of  $\theta_{23}=40^\circ,
42^\circ,45^\circ,48^\circ,50^\circ$ with
$\theta_{13}=0^\circ$ (first column), $5^\circ$ (second column), $7.5^\circ$
(third column) from neutrinos with NH  and $7.5^\circ$ (fourth 
column) from anti-neutrinos with IH.
}
\label{f:t13m68}
\end{figure*}

\begin{figure*}[htb]
\bc
\includegraphics[width=15.0cm,height=6cm,angle=0]{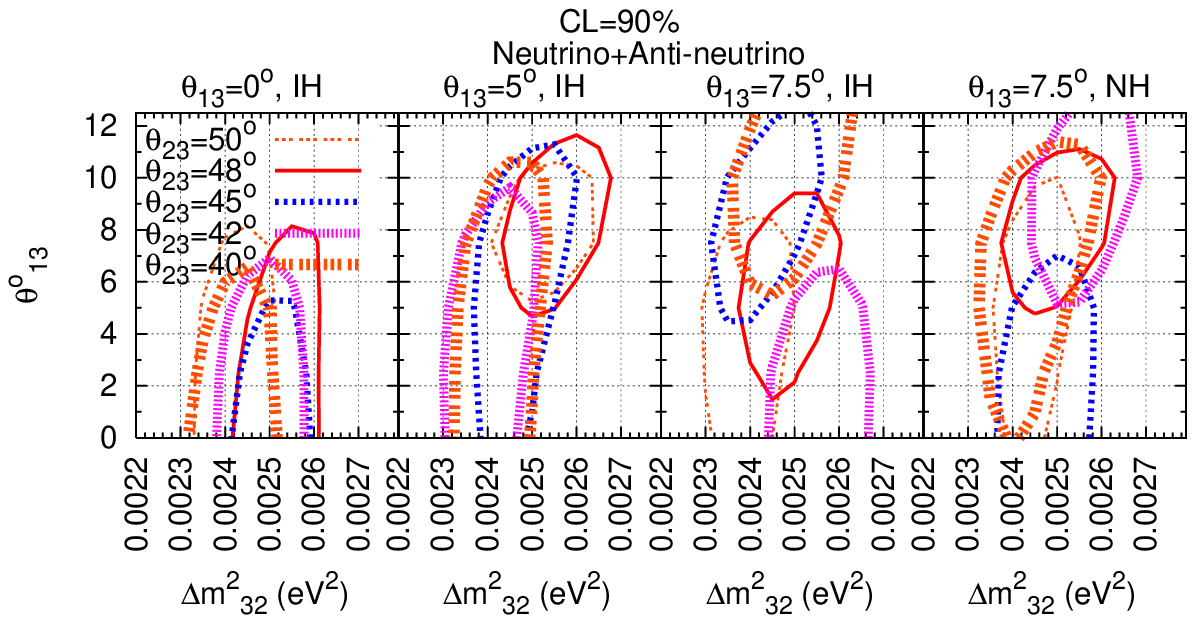}
\includegraphics[width=15.0cm,height=6cm,angle=0]{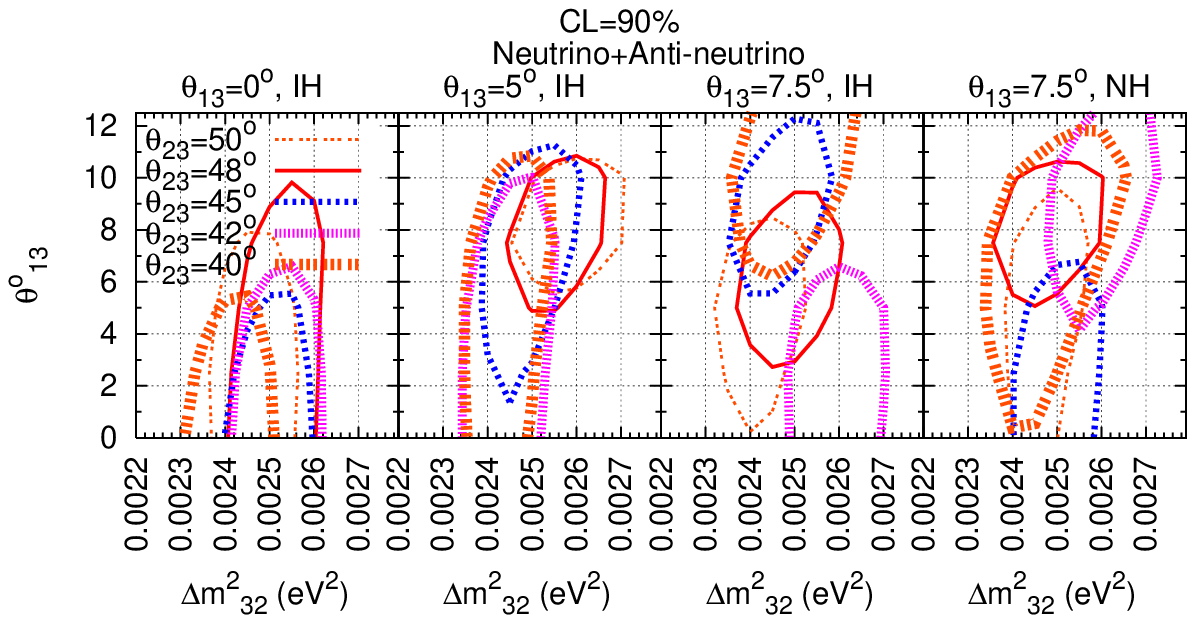}
\includegraphics[width=15.0cm,height=6cm,angle=0]{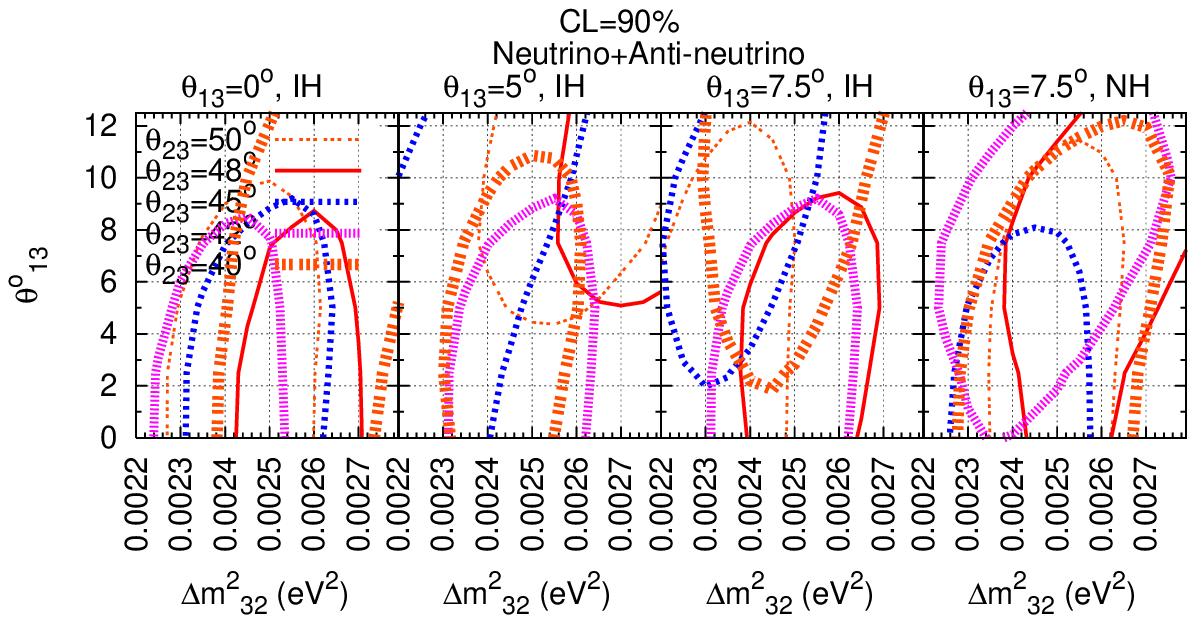}
\ec
\caption{\sf\small  
The same plots of fig. \ref{f:t13m68} but with 90\% CL.
}
\label{f:t13m90}
\end{figure*}

\begin{figure*}[htb]
\bc
\includegraphics[width=15.0cm,height=6cm,angle=0]{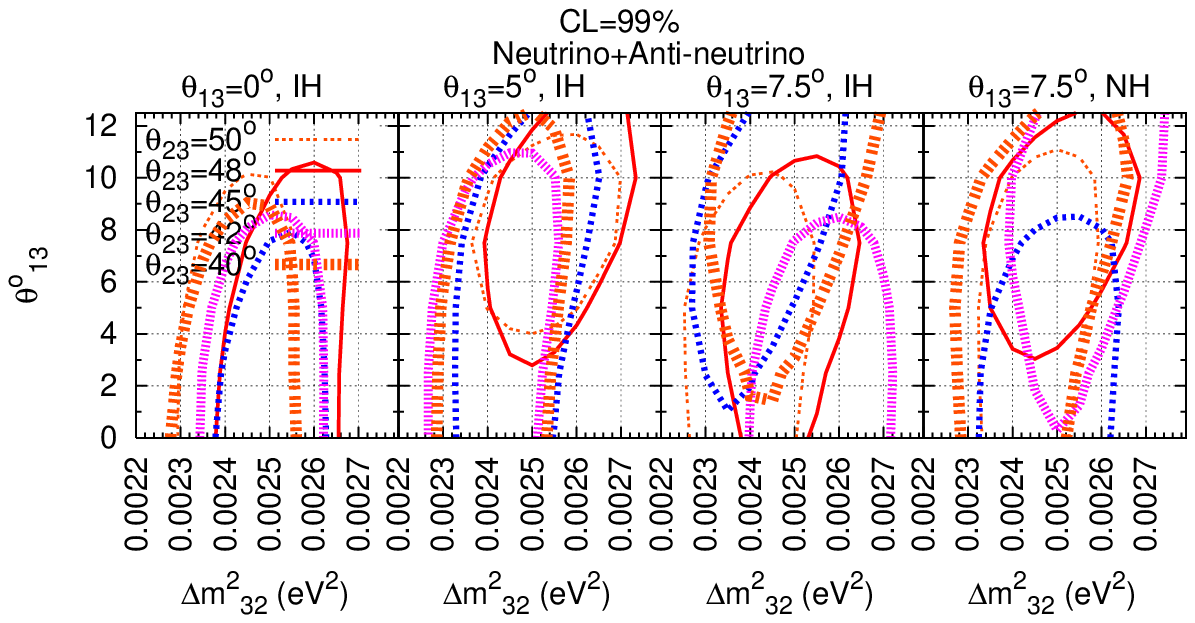}
\includegraphics[width=15.0cm,height=6cm,angle=0]{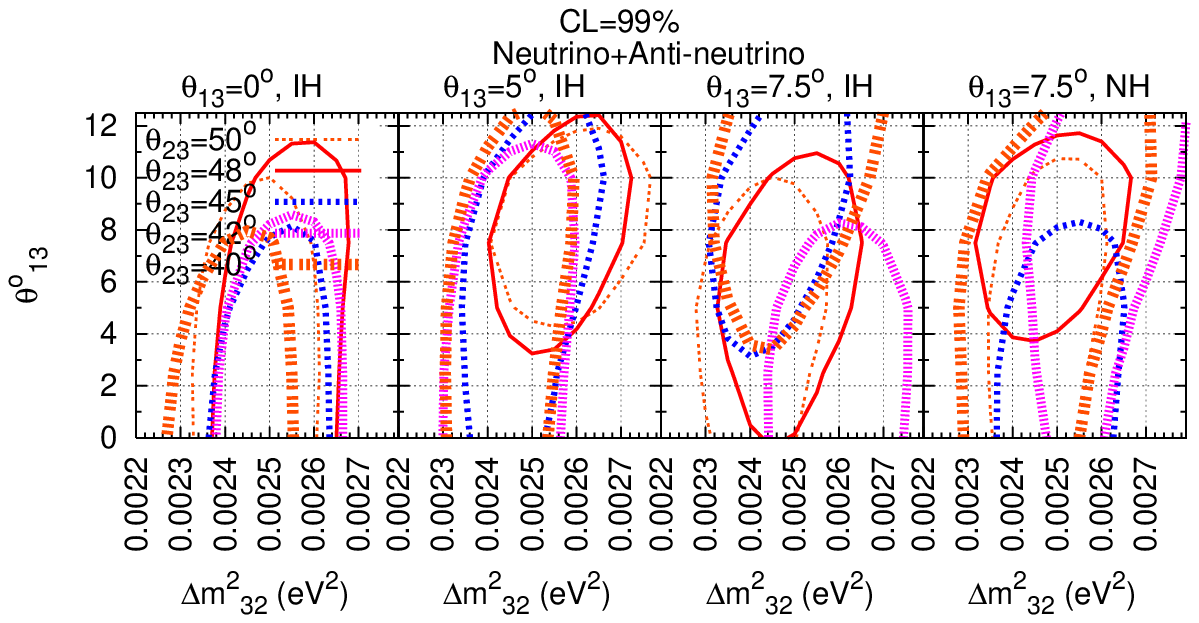}
\includegraphics[width=15.0cm,height=6cm,angle=0]{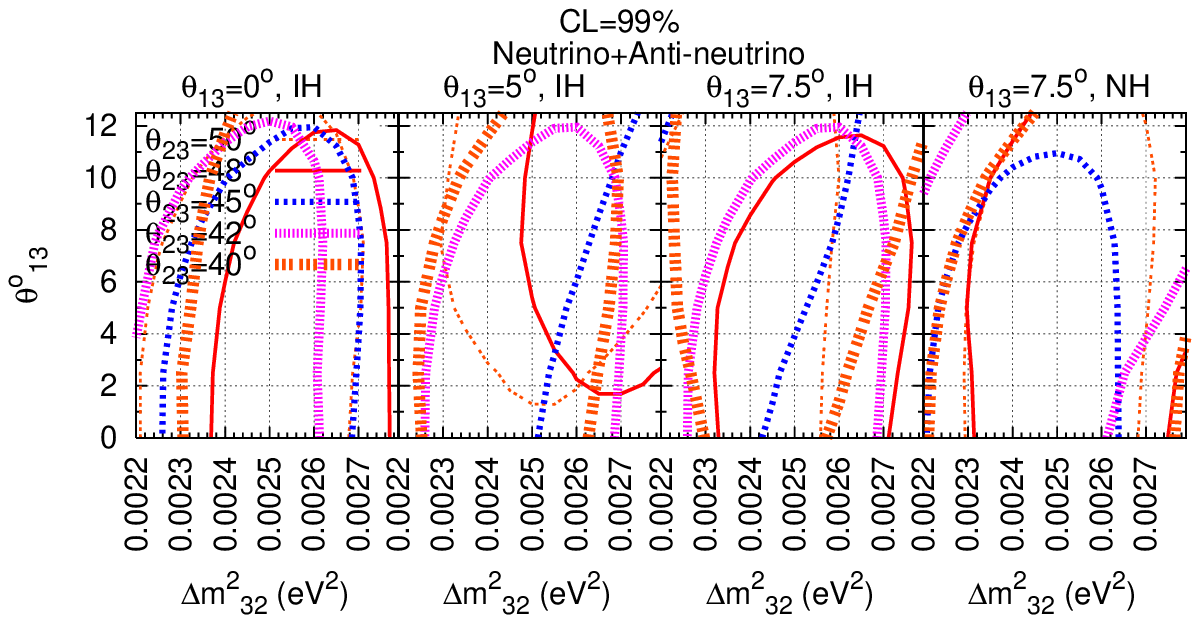}
\ec
\caption{\sf\small  
The same plots of fig. \ref{f:t13m68} but with 99\% CL.
}
\label{f:t13m99}
\end{figure*}

The difference in oscillation probability 
between two $\theta_{13}$ values for neutrinos with NH and for anti-neutrinos
with IH  becomes significant when matter effect comes in the picture 
(see eq. \ref{e:effective-mixing}). We have plotted the 
APS in $\theta_{13}-\Delta m_{32}^2$ plane considering both neutrinos 
and anti-neutrinos ({\it i.e.} with $\chi^2_{\rm total}=\chi^2_\nu+\chi^2_{\bar\nu}$)
%and with anti-neutrinos ({\it i.e.} with $\chi^2_{\bar\nu}$) for IH and with
%neutrinos  ({\it i.e.} with $\chi^2_{\nu}$)  for NH 
for different sets of input 
parameters at 68\%, 90\% and 99\% CL in fig \ref{f:t13m68},  \ref{f:t13m90}
and \ref{f:t13m99}, respectively for each type of binning of the data.
We see that the matter effect significantly constrains $\theta_{13}$ over
the present limit, which is a very stimulating result for atmospheric 
neutrino oscillation analysis.

\begin{figure*}[htb]
\bc
\includegraphics[width=15.0cm,height=6cm,angle=0]{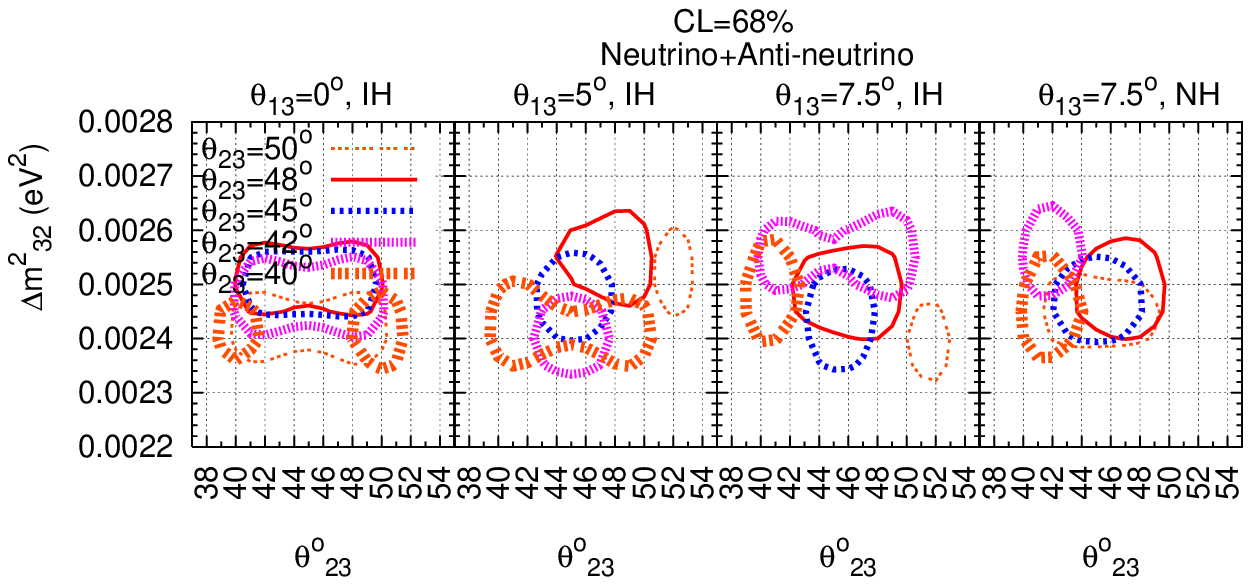}
\includegraphics[width=15.0cm,height=6cm,angle=0]{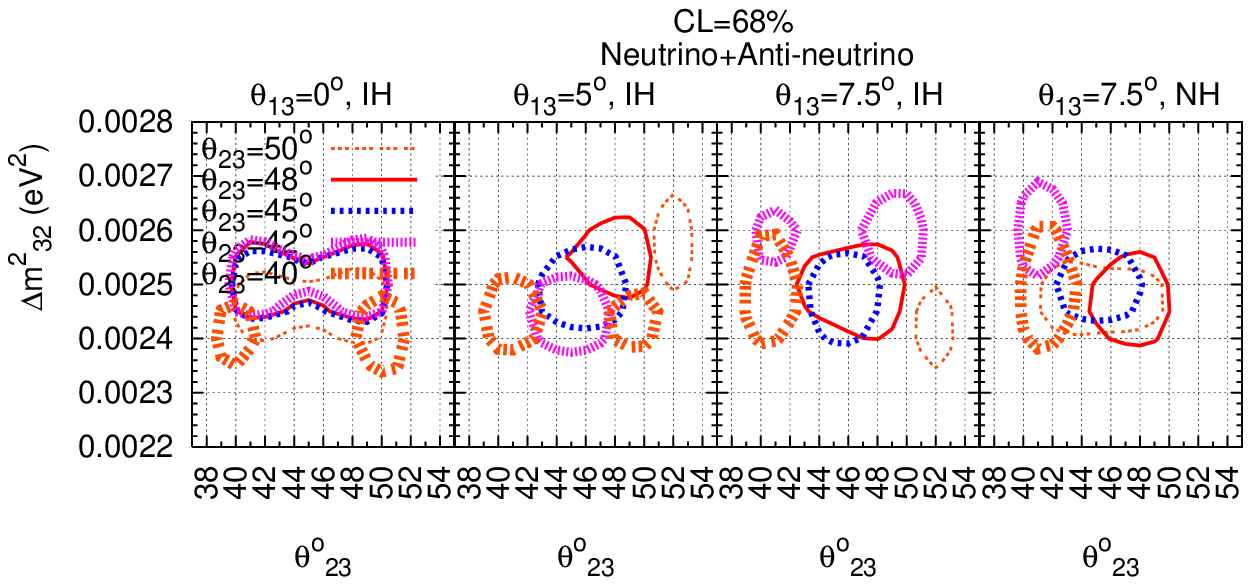}
\includegraphics[width=15.0cm,height=6cm,angle=0]{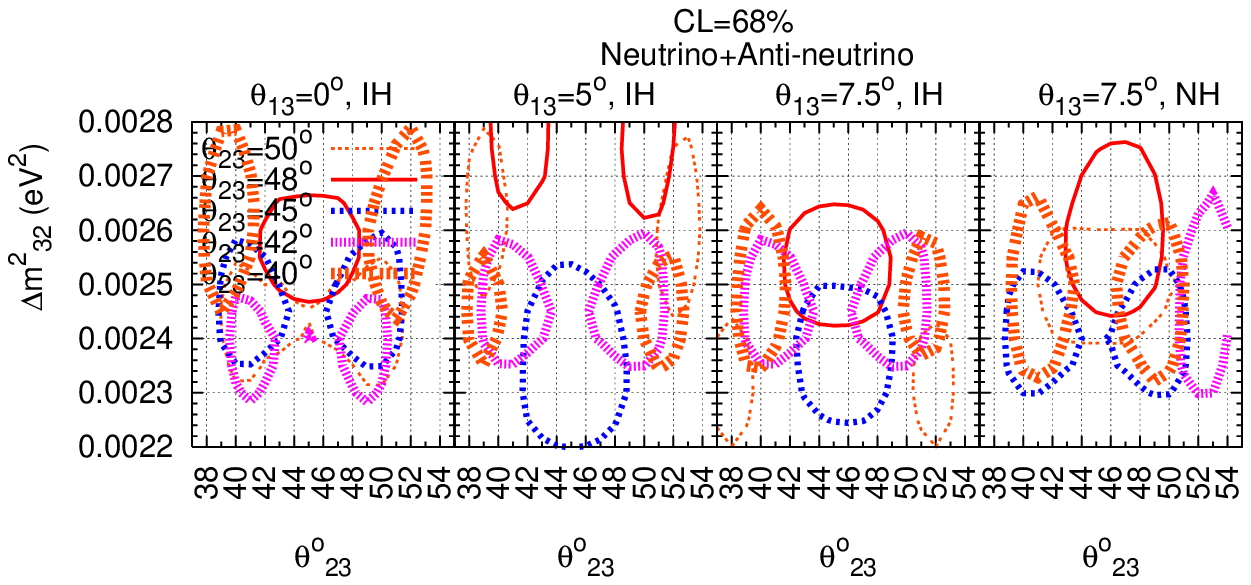}
\ec
\caption{\sf\small  
The 68\%  CL allowed regions in 
$\theta_{23}-\Delta m_{32}^2$ plane for  type I (top) type II (middle) and type III 
(bottom)  binning of the data  with the input of  $\theta_{23}=40^\circ,
42^\circ,45^\circ,48^\circ,50^\circ$ with
$\theta_{13}=0^\circ$ (first column), $5^\circ$ (second column), $7.5^\circ$
(third column) from neutrinos with NH  and $7.5^\circ$ (fourth 
column) from anti-neutrinos with IH.
}
\label{f:t23m68}
\end{figure*}

\begin{figure*}[htb]
\bc
\includegraphics[width=15.0cm,height=6cm,angle=0]{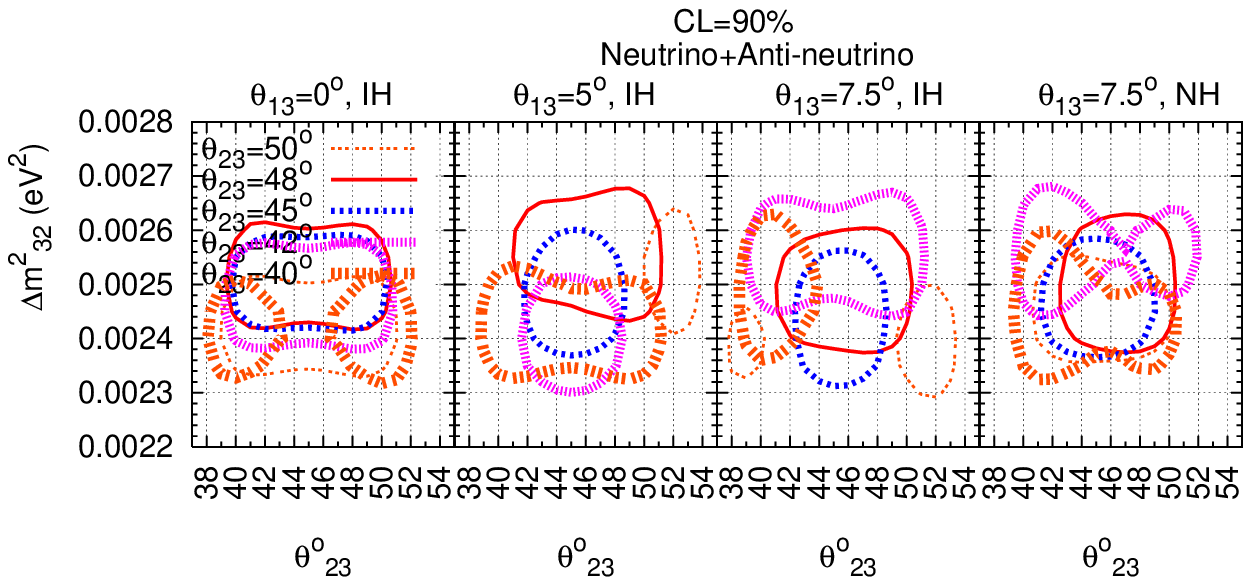}
\includegraphics[width=15.0cm,height=6cm,angle=0]{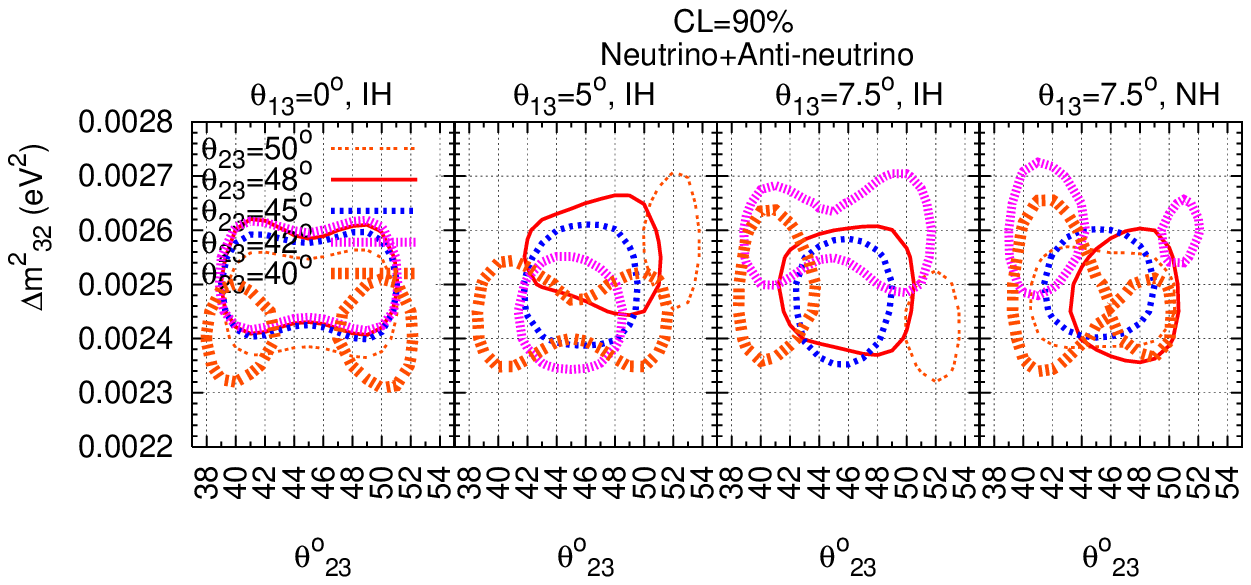}
\includegraphics[width=15.0cm,height=6cm,angle=0]{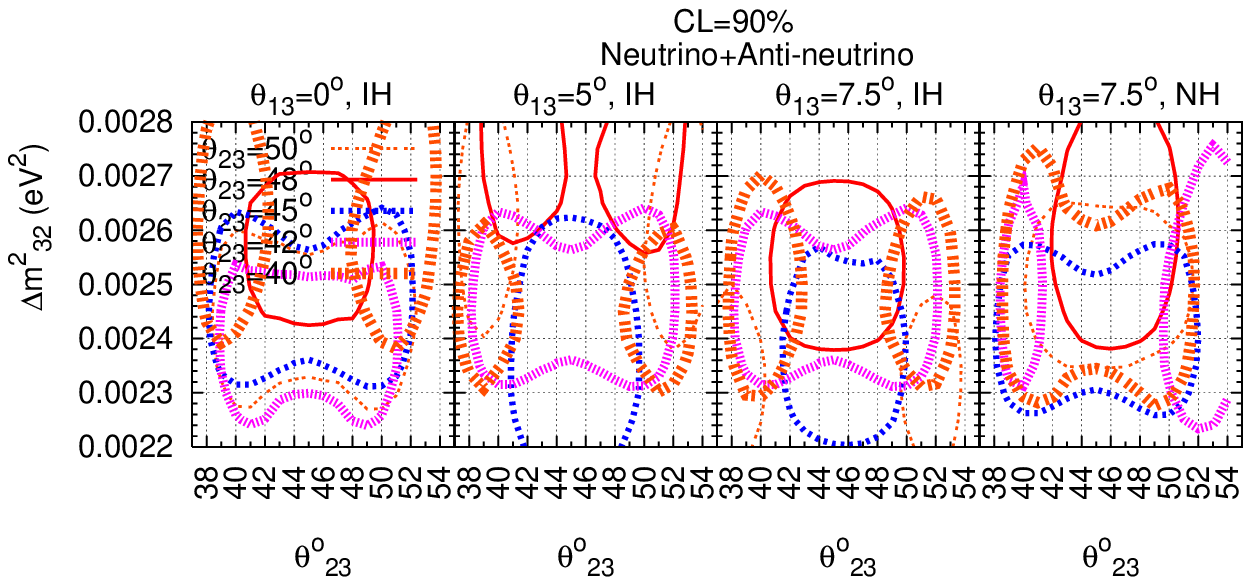}
\ec
\caption{\sf\small 
The same plots of fig. \ref{f:t23m68} but with 90\% CL.
 }
 \label{f:t23m90}
 \end{figure*}

\begin{figure*}[htb]
 \bc
 \includegraphics[width=15.0cm,height=6cm,angle=0]{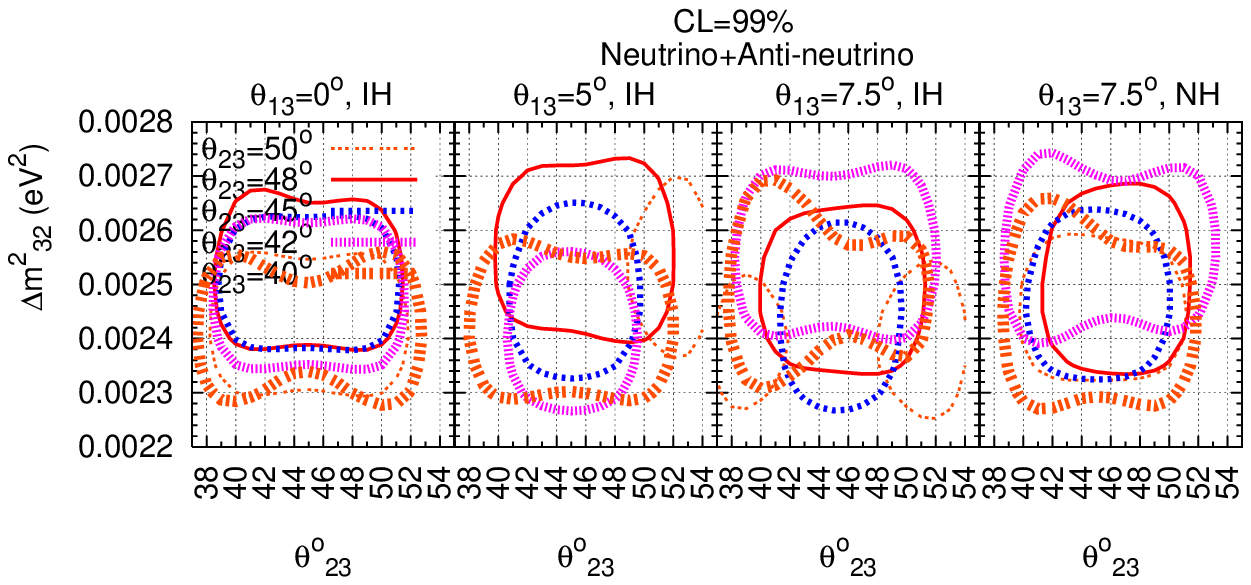}
 \includegraphics[width=15.0cm,height=6cm,angle=0]{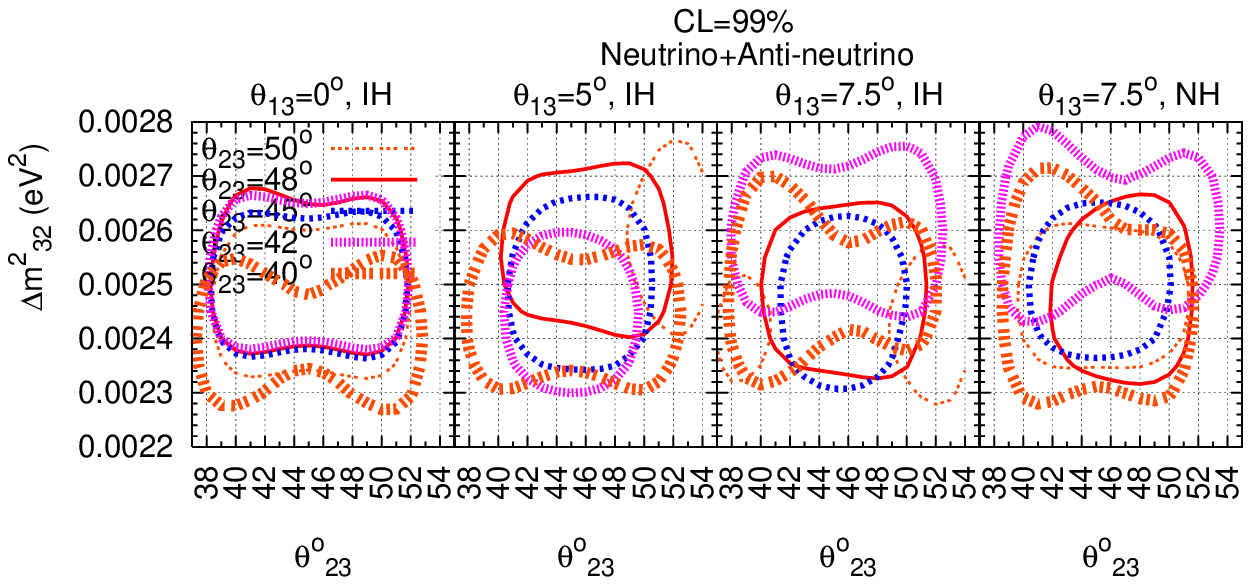}
\includegraphics[width=15.0cm,height=6cm,angle=0]{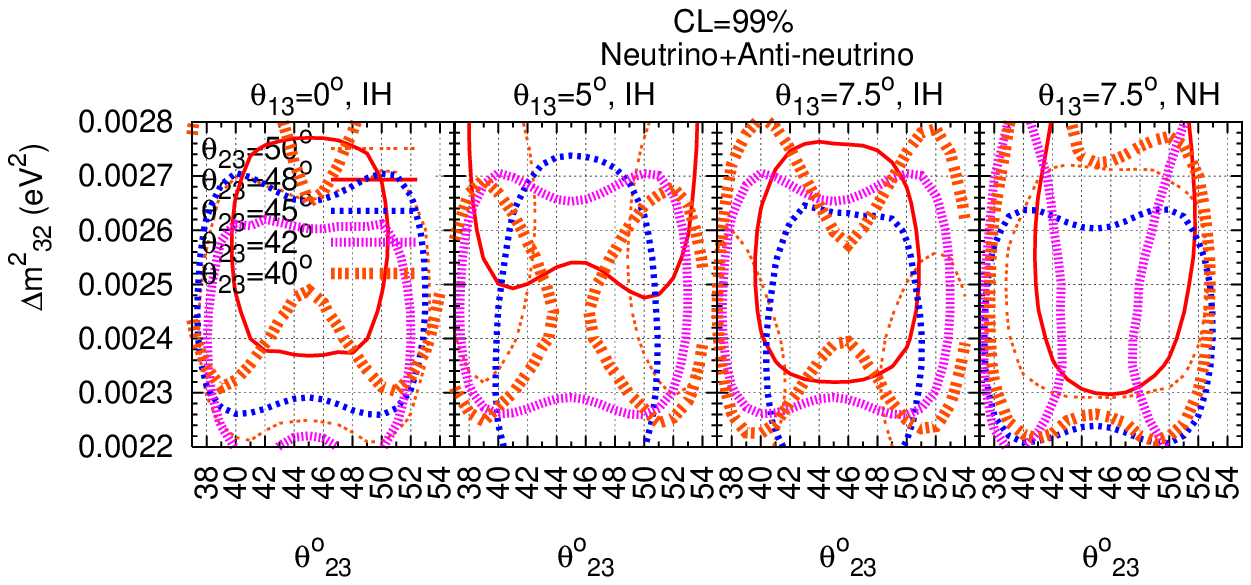}
 \ec
 \caption{\sf\small
The same plots of fig. \ref{f:t23m68} but with 99\% CL.
  }
  \label{f:t23m99}
  \end{figure*}

Again, for the APS in $\Delta m^2_{32}-\theta_{23}$ plane, $\theta_{13}$
is marginalized over the present allowed range. The APSs are shown 
in fig. \ref{f:t23m68},  \ref{f:t23m90} and  \ref{f:t23m99} at 
68\%, 90\% and 99\% CL, respectively for each type of binning of the data.
If the value of $\theta_{13}$ is nonzero, the matter effect plays a role 
in determination of  the octant of $\theta_{23}$ as discussed 
previously and also constrains the $\theta_{23}$ range (compare 
its range for zero and non-zero values of $\theta_{13}$). 
We find that for some combinations of ($\theta_{13}, ~ \theta_{23}$), 
the octant determination is possible. 

Now we will compare the APSs coming from different analysis method.
From the APSs it is clear that the $L/E$ analysis gives very poor results
compared to the other two methods. It happens due to the mixing of events from 
different $E$ and $L$ resolutions since the resolution widths are strongly
energy dependent. It should be noted here that we have not used any selection criteria
for the events, which might improve the results. 

Now we will compare the positive and the negative sides of the 
rest two methods.
We find a relatively stronger upper bound of $\Delta m^2_{32}$
in case of binning in the grids of $\log E - \log L$ plane than the other case. This is very
important since it comes from the events with high $E$ and low $L$ values.
The $L$ resolution is very poor at low $L$ and the statistics is low at high $E$. However, a
stronger bound is obtained for this special type of binning.  We will explain it with the
oscillation probability in vacuum, which is a sinusoidal function of $L/E$.
For a fixed $L$, the distance between two consecutive peaks in $E$
increases rapidly with $E$. Again, if we compare the distances between two
consecutive peaks in $E$ for two fixed values of $L$, it is larger for
smaller $L$ value. Therefore, as one goes to smaller $L$ values, this distance
in $E$ increases rapidly. So, these two consecutive peaks of the oscillation
in $E$ can be resolved with much better resolution as one goes gradually
from larger $L$ values to lower $L$ values. This is pictorially illustrated
in fig. \ref{f:p3d}. To get the reflection of this fact in $\chi^2$, the
finer binning at lower $L$ is essential.  Though the angular resolution is worsened
at lower $L$, but the rapid increase  of  $E$ resolution between two peaks
wins the competition here. This is the main advantage of this type of binning.
So, we binned the data in a two dimensional grids of $\log L-\log E$ plane
%\footnote{This is the first attempt of this type of binning. 
\footnote{This
captures the oscillation effect well in $\chi^2$ analysis without mixing events 
from different $E$ and $L$ resolutions.}.  
In type II this behavior is not 
taken into account in the binning of the data.  However, there is a disadvantage in 
type I that the bin size at high $L$ values is very large 
compared to type II, which gives weaker lower bound on $\Delta m_{32}^2$. 
So, the combination of type I and II (type I at the lower 
range of $L$ and type II for the rest) is a better choice than the individual cases. 
However, this is not studied in this paper, but is 
reflected when we compare two results.  
This is also demonstrated in terms of $\Delta \chi^2$  for a typical set of 
parameters in fig. \ref{f:dcm}. It should be noted here that 
the contrast between two methods would be prominent when the number of bins
in $L$ or $\cos\theta_{\rm zenith}$ will be relatively lowered than that used 
in this paper.

For a quantitative assessment of the result, we define the precision $P$ 
of a parameter $t$ as:  
\be P=2\left(\frac{t^{\rm max}-t^{\rm min}}{t^{\rm max}+t^{\rm min}}\right )\ee

We see, one can achieve the precision of 
$\Delta m_{32}^2 \approx$ $4.8-7.5\%~(5.4 - 8.0\%), ~6.9-10.9\%~(8.0-12.6\%)~{\rm and}~9.6-15.7\%~ (10.9-17.6\%)$, 
at 68\%, 90\% and 99\% CL, respectively in type I (II) method.
For the input with bi-maximal mixing of $\theta_{23}$, we find its precision 
in terms of $\sin^2\theta_{23}\approx $ $14.3-31.8\%~ (16.9-36.9\%),~ 21.6-36.8\% ~(22.4 - 41.7\%),{~ \rm and}~ 
 28.5-42.1\%~ (27.9-45.9\%)$ at 68\%, 90\%
and 99\% CL, respectively in type I (II) method.
The precision of $\theta_{13}$ is strongly dependent on its input value.
For $\theta_{13}=0$, we find its upper bound  $\approx 6.4^\circ~(8.0^\circ),~ 
8.0^\circ~(9.5^\circ)$ 
and $10.1^\circ~(11.5^\circ)$ at 68\%, 90\% and  99\% CL, respectively in type I (II) methods. 
The both lower and upper bounds are also possible for some 
combinations of ($\theta_{23}, \theta_{13}$) and it happens mainly 
for $\theta_{23}\gapp 45^\circ$.

\section{Conclusion}
In this paper we have binned the atmospheric data in three ways:
i) in the grids  of $\log E -\log L$ plane, ii) in the grids of
$\log E -\cos\theta_{\rm zenith}$ plane, and iii) in the bins of $\log (L/E)$.
We have performed a marginalized $\chi^2$ study over $\Delta m_{32}^2, ~\theta_{13}$
and $\theta_{23}$ for neutrinos and anti-neutrinos separately for each method.
Finally, we find that in spite of very poor resolutions at low $L$, which is the main
problem as $\Delta m_{32}^2$ goes to the upper range, one can  obtain a
relatively stronger upper bound in case of binning in $\log E-\log L$ plane
compared to the binning in $\log E-\cos\theta_{\rm zenith}$ plane. However, it is also found
from both analysis that  considerable precisions  of $\theta_{13}$ and $\Delta m_{32}^2$ 
can be achieved and the octant discrimination
can also be possible for some combinations of ($\theta_{23}, \theta_{13}$).



{\bf Acknowledgments:} I would like to acknowledge the excellent cluster computational 
facility of HRI, which makes the work possible. The research has been supported by the funds
from Neutrino Physics project at HRI.


\begin{thebibliography}{99}

\bibitem{Haines:1986yf}
  T.~J.~Haines {\it et al.},
  %``Calculation of Atmospheric Neutrino Induced Backgrounds in a Nucleon Decay
  %Search,''
  Phys.\ Rev.\ Lett.\  {\bf 57}, 1986 (1986).
  %%CITATION = PRLTA,57,1986;%%


%\bibitem{BeckerSzendy:1992hq}
%  R.~Becker-Szendy {\it et al.},
%  %``The Electron-neutrino and muon-neutrino content of the atmospheric flux,''
%  Phys.\ Rev.\  D {\bf 46}, 3720 (1992).
%  %%CITATION = PHRVA,D46,3720;%%

\bibitem{Hirata:1988uy}
  K.~S.~Hirata {\it et al.}  [KAMIOKANDE-II Collaboration],
  %``Experimental Study Of The Atmospheric Neutrino Flux,''
  Phys.\ Lett.\  B {\bf 205}, 416 (1988).
  %%CITATION = PHLTA,B205,416;%%

\bibitem{Fukuda:1998mi}
  Y.~Fukuda {\it et al.}  [Super-Kamiokande Collaboration],
  %``Evidence for oscillation of atmospheric neutrinos,''
  Phys.\ Rev.\ Lett.\  {\bf 81}, 1562 (1998)
  [arXiv:hep-ex/9807003].
  %%CITATION = PRLTA,81,1562;%%

\bibitem{Fogli:2008ig}
  G.~L.~Fogli {\it et al.},
  %``Observables sensitive to absolute neutrino masses (Addendum),''
  Phys.\ Rev.\  D {\bf 78}, 033010 (2008)
  [arXiv:0805.2517 [hep-ph]].
  %%CITATION = PHRVA,D78,033010;%%

\bibitem{Arumugam:2005nt}
  V.~Arumugam {\it et al.}  [INO Collaboration], INO-2005-01.
  %``India-based Neutrino Observatory: Interim project report. Vol. 1,''
  %%CITATION = INO-2005-01;%%



\bibitem{Jung:1999jq}
  C.~K.~Jung,
  %``Feasibility of a next generation underground water Cherenkov detector:
  %UNO,''
  AIP Conf.\ Proc.\  {\bf 533}, 29 (2000)
  [arXiv:hep-ex/0005046].
  %%CITATION = APCPC,533,29;%%

\bibitem{Itow:2001ee}
  Y.~Itow {\it et al.}  [The T2K Collaboration],
  %``The JHF-Kamioka neutrino project,''
  arXiv:hep-ex/0106019.
  %%CITATION = HEP-EX/0106019;%%

\bibitem{Ayres:2004js}
  D.~S.~Ayres {\it et al.}  [NOvA Collaboration],
  %``NOvA proposal to build a 30-kiloton off-axis detector to study neutrino
  %oscillations in the Fermilab NuMI beamline,''
  arXiv:hep-ex/0503053.
  %%CITATION = HEP-EX/0503053;%%

\bibitem{Nakamura:2003hk}
  K.~Nakamura,
  %``Hyper-Kamiokande: A next generation water Cherenkov detector,''
  Int.\ J.\ Mod.\ Phys.\  A {\bf 18}, 4053 (2003).
  %%CITATION = IMPAE,A18,4053;%%

\bibitem{GonzalezGarcia:2007ib}
  See the review, M.~C.~Gonzalez-Garcia and M.~Maltoni,
  %``Phenomenology with Massive Neutrinos,''
  Phys.\ Rept.\  {\bf 460}, 1 (2008)
  [arXiv:0704.1800 [hep-ph]] and the references there in.
  %%CITATION = PRPLC,460,1;%%

\bibitem{Choubey:2005zy}
  S.~Choubey and P.~Roy,
  %``Probing the deviation from maximal mixing of atmospheric neutrinos,''
  Phys.\ Rev.\  D {\bf 73}, 013006 (2006)
  [arXiv:hep-ph/0509197].
  %%CITATION = PHRVA,D73,013006;%%


\bibitem{Gandhi:2007td}
  R.~Gandhi, P.~Ghoshal, S.~Goswami, P.~Mehta, S.~U.~Sankar and S.~Shalgar,
  %``Mass Hierarchy Determination via future Atmospheric Neutrino Detectors,''
  Phys.\ Rev.\  D {\bf 76}, 073012 (2007)
  [arXiv:0707.1723 [hep-ph]].
  %%CITATION = PHRVA,D76,073012;%%

\bibitem{Indumathi:2004kd}
  D.~Indumathi and M.~V.~N.~Murthy,
  %``A question of hierarchy: Matter effects with atmospheric neutrinos and
  %anti-neutrinos,''
  Phys.\ Rev.\  D {\bf 71}, 013001 (2005)
  [arXiv:hep-ph/0407336].
  %%CITATION = PHRVA,D71,013001;%%

\bibitem{Petcov:2005rv}
  S.~T.~Petcov and T.~Schwetz,
     %``Determining the neutrino mass hierarchy with atmospheric neutrinos,''
       Nucl.\ Phys.\  B {\bf 740}, 1 (2006)
        [arXiv:hep-ph/0511277].
	          %%CITATION = NUPHA,B740,1;%%


\bibitem{Indumathi:2006gr}
  D.~Indumathi, M.~V.~N.~Murthy, G.~Rajasekaran and N.~Sinha,
  %``Neutrino oscillation probabilities: Sensitivity to parameters,''
  Phys.\ Rev.\  D {\bf 74}, 053004 (2006)
  [arXiv:hep-ph/0603264].
  %%CITATION = PHRVA,D74,053004;%%

\bibitem{Agarwalla:2005we}
  S.~K.~Agarwalla, A.~Raychaudhuri and A.~Samanta,
  %``Exploration prospects of a long baseline beta beam neutrino experiment
  %with an iron calorimeter detector,''
  Phys.\ Lett.\  B {\bf 629}, 33 (2005)
  [arXiv:hep-ph/0505015].
  %%CITATION = PHLTA,B629,33;%%

\bibitem{Agarwalla:2006vf}
  S.~K.~Agarwalla, S.~Choubey and A.~Raychaudhuri,
  %``Neutrino mass hierarchy and theta(13) with a magic baseline beta-beam
  %experiment,''
  Nucl.\ Phys.\  B {\bf 771}, 1 (2007)
  [arXiv:hep-ph/0610333].
  %%CITATION = NUPHA,B771,1;%%


\bibitem{Agarwalla:2008gf}
  S.~K.~Agarwalla, S.~Choubey, A.~Raychaudhuri and W.~Winter,
  %``Optimizing the greenfield Beta-beam,''
  JHEP {\bf 0806}, 090 (2008)
  [arXiv:0802.3621 [hep-ex]].
  %%CITATION = JHEPA,0806,090;%%


      \bibitem{Datta:2003dg}
        A.~Datta, R.~Gandhi, P.~Mehta and S.~Uma Sankar,
        %``Atmospheric neutrinos as a probe of CPT and Lorentz violation,''
        Phys.\ Lett.\  B {\bf 597}, 356 (2004)
        [arXiv:hep-ph/0312027].
        %%CITATION = PHLTA,B597,356;%%

\bibitem{inopaper}
For more details see, http://www.imsc.res.in/~ino/Talks/papers.html

\bibitem{Ashie:2005ik}
  Y.~Ashie {\it et al.}  [Super-Kamiokande Collaboration],
  %``A Measurement of Atmospheric Neutrino Oscillation Parameters by
  %Super-Kamiokande I,''
  Phys.\ Rev.\  D {\bf 71}, 112005 (2005)
  [arXiv:hep-ex/0501064].
  %%CITATION = PHRVA,D71,112005;%%

\bibitem{Ashie:2004mr}
  Y.~Ashie {\it et al.}  [Super-Kamiokande Collaboration],
  %``Evidence for an oscillatory signature in atmospheric neutrino
  %oscillation,''
  Phys.\ Rev.\ Lett.\  {\bf 93}, 101801 (2004)
  [arXiv:hep-ex/0404034].
  %%CITATION = PRLTA,93,101801;%%


\bibitem{Mikheev:1986gs}
  S.~P.~Mikheev and A.~Y.~Smirnov,
  %``Resonance enhancement of oscillations in matter and solar neutrino
  %spectroscopy,''
  Sov.\ J.\ Nucl.\ Phys.\  {\bf 42}, 913 (1985)
  [Yad.\ Fiz.\  {\bf 42}, 1441 (1985)];
  %%CITATION = YAFIA,42,1441;%%
%\bibitem{Mikheev:1986wj}
%  S.~P.~Mikheev and A.~Y.~Smirnov,
  %``Resonant amplification of neutrino oscillations in matter and solar
  %neutrino spectroscopy,''
  Nuovo Cim.\  C {\bf 9}, 17 (1986).
  %%CITATION = NUCIA,9C,17;%%

\bibitem{Wolfenstein:1977ue}
  L.~Wolfenstein,
  %``Neutrino oscillations in matter,''
  Phys.\ Rev.\  D {\bf 17}, 2369 (1978).
  %%CITATION = PHRVA,D17,2369;%%

\bibitem{Giunti:1997fx}
  C.~Giunti, C.~W.~Kim and M.~Monteno,
  %``Atmospheric neutrino oscillations with three neutrinos and a mass
  %hierarchy,''
  Nucl.\ Phys.\  B {\bf 521}, 3 (1998)
  [arXiv:hep-ph/9709439].
  %%CITATION = NUPHA,B521,3;%%

\bibitem{Samanta:2006sj}
  A.~Samanta,
  %``The mass hierarchy with atmospheric neutrinos at INO,''
  arXiv:hep-ph/0610196.
  %%CITATION = HEP-PH/0610196;%%

\bibitem{Casper:2002sd}
  D.~Casper,
  %``The nuance neutrino physics simulation, and the future,''
  Nucl.\ Phys.\ Proc.\ Suppl.\  {\bf 112}, 161 (2002)
  [arXiv:hep-ph/0208030].
  %%CITATION = NUPHZ,112,161;%%

\bibitem{Honda:2004yz}
M.~Honda, T.~Kajita, K.~Kasahara and S.~Midorikawa,
%``A new calculation of the atmospheric neutrino flux in a 3-dimensional
%scheme,''
Phys.\ Rev.\ D {\bf 70}, 043008 (2004)
[arXiv:astro-ph/0404457].
%%CITATION = ASTRO-PH 0404457;%%


%\bibitem{Fukuda:1994mc}
%  Y.~Fukuda {\it et al.}  [Kamiokande Collaboration],
%  %``Atmospheric muon-neutrino / electron-neutrino ratio in the multiGeV energy
%  %range,''
%  Phys.\ Lett.\  B {\bf 335}, 237 (1994).
%  %%CITATION = PHLTA,B335,237;%%


\bibitem{Fogli:2002pt}
  G.~L.~Fogli, E.~Lisi, A.~Marrone, D.~Montanino and A.~Palazzo,
  %``Getting the most from the statistical analysis of solar neutrino
  %oscillations,''
  Phys.\ Rev.\  D {\bf 66}, 053010 (2002)
  [arXiv:hep-ph/0206162].
  %%CITATION = PHRVA,D66,053010;%%


\bibitem{Apollonio:1999ae}
M.~Apollonio {\it et al.}  [CHOOZ Collaboration],
%``Limits on Neutrino Oscillations from the CHOOZ Experiment,''
Phys.\ Lett.\  B {\bf 466}, 415 (1999)
[arXiv:hep-ex/9907037].

\end{thebibliography}
\end{document}